\documentclass[times,twocolumn,final]{elsarticle}

\usepackage{medima}
\usepackage{framed,multirow}

\usepackage{amssymb}
\usepackage{latexsym}

\usepackage{url}
\usepackage{xcolor}

\usepackage{hyperref}

\usepackage{xtab,afterpage}
\usepackage{lineno}
\usepackage{color}
\usepackage{amssymb}
\usepackage{amsmath}
\usepackage{graphicx}
\usepackage{caption}
\usepackage{tabularx}
\usepackage{url}
\usepackage{makecell}
\usepackage{multirow}
\usepackage{appendix}
\usepackage{comment}
\usepackage{lscape} 
\usepackage{tablefootnote}
\usepackage{bm}
\usepackage[ruled, vlined]{algorithm2e}
\usepackage{caption}
\usepackage{subcaption}

\DeclareMathOperator*{\argmax}{argmax}

\newcommand{\jei}[1]{\color{blue}{Comentario de Eugenio: #1}\normalcolor}

\definecolor{newcolor}{rgb}{.8,.349,.1}

\journal{Medical Image Analysis}

\begin{document}

\verso{A. Casamitjana \textit{et~al.}}

\begin{frontmatter}

\title{USLR: an open-source tool for unbiased and smooth longitudinal registration of brain MRI
}

\author[1,2,8]{Adrià \snm{Casamitjana}\corref{cor1}}
\cortext[cor1]{Corresponding author: 
  Email: a.casamitjana@ub.edu}
\author[2,3,4]{Roser \snm{Sala-Llonch}
}
\author[5]{Karim \snm{Lekadir}
}
\author[6,7, 8]{Juan Eugenio \snm{Iglesias}
}

\address[1]{Universitat Politècnica de Catalunya. Barcelona, 08034, Spain}
\address[2]{Institut de Neurociències Department of Biomedicine, Faculty of Medicine, University of Barcelona, Barcelona, 08036, Spain}
\address[3]{Institut d’Investigacions Biomèdiques August Pi i Sunyer (IDIBAPS). Barcelona, 08036, Spain}
\address[4]{Centro de Investigación Biomédica en Red de Bioingeniería, Biomateriales y Nanomedicina (CIBER-BBN), Barcelona, 08036, Spain.}
\address[5]{Departament de Matemàtiques i Informàtica, Universitat de Barcelona, Artificial Intelligence in Medicine Lab (BCN-AIM), Barcelona, Spain}
\address[6]{Martinos Center for Biomedical Imaging, Massachusetts General Hospital and Harvard Medical School}
\address[7]{Computer Science and Artificial Intelligence Laboratory, Massachusetts Institute of Technology}
\address[8]{Centre for Medical Image Computing, University College London}

\received{N/A}
\finalform{N/A}
\accepted{N/A}
\availableonline{N/A}
\communicated{N/A}

\begin{abstract}
We present USLR, a computational framework for longitudinal registration of brain MRI scans to estimate nonlinear image trajectories that are smooth across time, unbiased to any timepoint, and robust to imaging  artefacts. It operates on the Lie algebra parameterisation of spatial transforms (which is compatible with rigid transforms and stationary velocity fields for nonlinear deformation) and takes advantage of log-domain properties to solve the problem using Bayesian inference. USRL estimates rigid and nonlinear registrations that: (\emph{i}) bring all timepoints to an unbiased subject-specific space; and  (\emph{ii}) compute a smooth trajectory across  the imaging time-series.  We capitalise on learning-based registration algorithms and closed-form expressions for fast inference. A use-case Alzheimer's disease study is used to showcase the benefits of the pipeline in multiple fronts, such as time-consistent image segmentation to reduce intra-subject variability, subject-specific prediction or population analysis using tensor-based morphometry. We demonstrate that such approach improves upon cross-sectional methods in identifying group differences, which can be helpful in detecting more subtle atrophy levels or in reducing sample sizes in clinical trials. The code is publicly available in \url{https://github.com/acasamitjana/uslr}
\end{abstract}


\end{frontmatter}


\section{Introduction}
\label{sec:intro}


Many of the central themes in neuroimaging, such as the effect of ageing, disease progression, or the effectiveness of a treatment, are intrinsically longitudinal. 
While cross-sectional studies are restricted to measuring population trajectories that may hide the true evolution of a given biomarker \citep{nyberg2010longitudinal}, longitudinal analysis uncovers truly individual trajectories that reduce confounding effects and dataset bias and results in better estimates -- even in situations where population averages are aligned with the true effect \citep{maxwell2007bias,kraemer2000can}. This increased power of longitudinal studies presents several opportunities, including: \textit{(i)}~better sensitivity and specificity, that could be used to detect different, partially overlapping atrophy patterns; \textit{(ii)}~reduced sample sizes for a target effect size; and \textit{(iii)}~new surrogate endpoints for therapeutic interventions. Most importantly, longitudinal analysis produces individualised measures that are useful for a wealth of applications, such as post-treatment followup or monitoring of disease progression.

There exists a myriad of statistical models for data analysis that can appropriately handle longitudinal data, such as repeated measures ANOVA, linear mixed effects regression, or growth models, among others \citep{garcia2017statistical}. In the context of medical imaging, it is crucial to carefully design an image processing pipeline (e.g., spatial normalisation or segmentation) prior to feeding such statistical models with the appropriate data. However, most state-of-the-art brain image processing techniques are developed for cross-sectional settings and suffer from poor measurement reliability -- which is a common limiting factor in longitudinal studies \cite{morey2010scan,karch2019identifying}. Instead, longitudinal processing techniques are able to capture time dynamics and capitalise on the redundancy in the available repeated measures to produce a more consistent and reliable result. For example, subject-specific templates  represent the ``average'' anatomy of a subject through time and can be used for longitudinal brain segmentation \citep{iglesias2016bayesian,cerri2023open}. Similarly, groupwise registration techniques can be used for more reliable spatial normalisation to a common subject-specific template~\citep{joshi2004unbiased, reuter2012within}. In conclusion, using an appropriate longitudinal processing followed by longitudinal statistical methods (e.g., LME or repeated measures ANOVA) might increase trustworthiness of the results.

Nonetheless, longitudinal processing streams need to take into account several considerations. Primarily, uncorrelated sources of variability between timepoints, such as intensity inhomogeneities, subject motion and the appearance and evolution of brain lesions. All these are identified as major causes of error in atrophy estimation methods~\citep{sharma2013estimation}. Likewise, longitudinal processing should be robust against intensity changes due to updates on the sequence, machine or scanning site during the follow up period as well as the inclusion of different MRI sequences, scanners, resolutions and field strengths between participants in multi-site studies~\citep{lee2019estimating}. Recent works on domain randomisation are designed to tackle these issues in several applications, such as for image super-resolution~\citep{iglesias2023synthsr}, segmentation~\citep{billot2023synthseg} and registration~\citep{hoffmann2022synthmorph}. 
Additionally, different sources of bias, such as interpolation asymmetries (i.e., using a given population template or choosing a timepoint as reference image) or excessive temporal regularisation, could hinder the power of the method or lead to false findings \citep{reuter2012within,yushkevich2010bias}.

The vast majority of classical existing registration algorithms, based on numerical optimisation, are designed for pairwise alignment. Some examples are those implemented in widespread registration packages like NiftyReg \citep{modat2010fast}, ANTs \citep{avants2008symmetric}, Elastix \citep{klein2009elastix}, DARTEL \citep{ashburner2007fast}, FNIRT~\citep{andersson2007non} or IRTK~\citep{schnabel2001generic}.
Moreover, amidst the deep learning revolution, multiple learning based registration methods emerged both in supervised \citep{yang2017quicksilver,young2022superwarp} and unsupervised settings \citep{de2019deep}. Again, most of these methods are trained using cross-sectional data. For example, the widespread Voxelmorph framework predicts a dense deformation field that registers pairs of T1w images independently \citep{balakrishnan2019voxelmorph}. Its extension presented in \cite{hoffmann2022synthmorph} is able to register pairs of images of any contrast thus handling differences in scanners, sequences and protocols. Several frameworks for joint linear and nonlinear learning-based registration have been introduced, such as DLIR \cite{de2019deep} or, more recently, EasyReg \citep{iglesias2023ready} 

Many longitudinal studies are limited to 2 timepoints so that cross-sectional registration pipelines could be used. For example, in lesion follow-up studies, most often the baseline image is considered as reference as it usually contains smaller lesions~\citep{diez2014intensity,dufresne2020joint}. The work in \cite{diez2014intensity} compare different linear and nonlinear registration methods to quantify the evolution of multiple sclerosis (MS) lesions. Seemingly, the authors in \cite{dufresne2020joint} propose a joint model for registration and change detection in MS. In the context of radiotherapy, post treatment images are typically registered to the pre-treatment image for therapy follow-up \cite{lee2021deformation, lee2023seq2morph}. Lately, in the field of brain tumour resection \cite{baheti2021brain}, several learning based approaches have been introduced, encouraging inverse consistency in the loss function during training \citep{wodzinski2022unsupervised, mok2022unsupervised} and handling missing correspondences between images \citep{mok2022robust}. Here, baseline and follow-up images are used as reference interchangeably.

Notably, available longitudinal registration pipelines have limitations in some way or another; for instance, FreeSurfer is restricted to using rigid transforms. Other existing methods, such as NiftyReg and ANTs, can be used for groupwise registration by alternating between mean template computation and pairwise registration to such template. For example, the iterative approach presented in \citep{aubert2013new} is initialised - thus biased - with the standard ICBM152 template. Moreover, registering each timepoint independently to the template results in jagged longitudinal trajectories. A similar work to ours is presented in \cite{hadj2016longitudinal}. Their method adopt the baseline image as subject-specific template. Besides, they use a classic model-based registration algorithm \citep{lorenzi2013lcc} which is typically more computationally demanding than learning-based methods. Unlike our work, the computed trajectories lie on MNI space instead of subject space and they do not provide time-consistent segmentations. In \cite{agier2020hubless}, the authors used the same graph structure as our observational graph in the context of groupwise registration of multiple subjects. This structure is very time and memory consuming as it needs pairwise registration of all available timepoints using slow model-based algorithms. To reduce computational demands, they scale down image dimensionality by working on image keypoints. In our framework we benefit from learning based registration algorithms to deal with the large computational demands of longitudinal registration.

The contributions of this work are threefold:
\begin{itemize}
    \item First, we present a theoretical framework for longitudinal registration of MRI scans that is explicitly smooth along time, unbiased to any timepoint and robust to irregular follow up times and varying number of timepoints. This method is based on Lie algebra parameterisation of spatial transforms. Thanks to our choice of registration algorithms, the framework is invariant to the MRI sequence, contrast and resolution. 
    \item Second, we apply the framework to find subject-specific MRI trajectories that are continuous across time as well as a subject-specific template. For this purpose, we describe two different models of spatial transforms that are applied sequentially: a rigid transform and a nonlinear diffeomorphism based on stationary velocity fields. 
    \item Third, we use subject-specific longitudinal deformations to compute time consistent segmentations for all timepoints from initial cross-sectional segmentations using a label fusion approach. 
\end{itemize}

The rest of this manuscript is organised as follows: in Section~\ref{sec:uslr_framework}, we describe the USLR framework and thoroughly discuss the probabilistic model, the benefits of using Lie algebra parameterisations, the inference method, and the algorithms used for rigid and nonlinear registration. In Section~\ref{sec:pipeline}, we present two methods that build on USLR to compute  (\emph{i}) a single stationary subject-specific trajectory and (\emph{ii}) a time-consistent segmentation.  In Section~\ref{sec:results}, we  validate and demonstrate the benefits of the USLR framework, illustrated in a use-case group study. Finally, Section~\ref{sec:discussion} summarises the main advantages of USLR as well as discusses the limitations and future directions of this work.

\section{USLR framework}
\label{sec:uslr_framework}

Our unbiased, smooth, longitudinal registration algorithm (``USLR'') builds on a probabilistic model of joint diffeomorphic deformations that we first presented in the context of 3D histology reconstruction \citep{casamitjana2022robust}. Here, we extend this methodology to impose smoothness and consistency constraints on  longitudinal deformations connecting the MRI scans of a subject at different time points. The USLR framework can be applied indistinctly to rigid and non-rigid transforms as long as the latter are parameterised using stationary velocity fields (SVFs) - thus being compatible with multiple registration algorithms.

USLR uses rigid transforms to model changes on the orientation (rotation) and position (translation) of images along the longitudinal course of each participant. Subsequently, non-rigid deformations are used to model local changes in brain tissue configuration, such as changes related to ageing or neurodegeneration. Moreover, we use domain randomisation techniques \citep{billot2023synthseg,billot2023robust,hoffmann2022synthmorph} to deal with acquisition MRI artefacts and differences in scanning platforms and pulse sequences (resolution, MR contrast).

In the rest of this section, we describe the USLR probabilistic model and the  parameterisation of rigid and nonlinear transforms in the log-space that is required by the USLR inference algorithm -- which is  subsequently presented.

\subsection{Preliminaries: graph representation}
\label{sec:slr_preliminaries}
\begin{figure}[h]
    \centering
    \includegraphics[width=0.8\columnwidth]{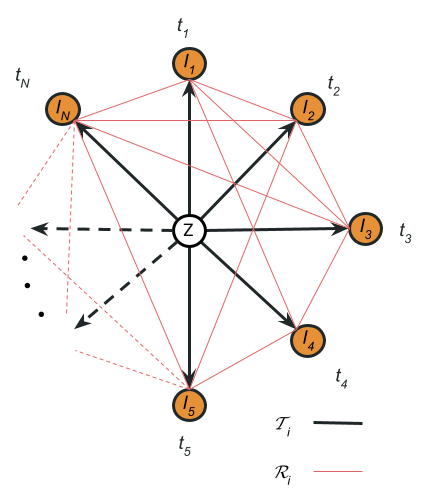}
    \caption{The graph structure $\mathcal{G}$, where all timepoints ``orbit'' around the unobserved template. In black, our choice of spanning tree of the graph, where all timepoints are connected through the template. The direction of the associated deformation fields is from the template to the timepoints, as indicated by the arrows In red, we draw the observational graph describing the dense pairwise ``noisy'' registrations of all timepoints. The direction of this transforms is arbitrary for each subject and known throughout the pipeline.}
    \label{fig:spanning-tree}
\end{figure}

Let us consider a given subject with a set of $N$ longitudinal MRI scans denoted as $I_1, I_2, \cdots, I_{N}$ acquired at time points $t_1, t_2, \cdots, t_{N}$, respectively. There is no assumption on the spacing between time points (which does not need to be uniform) nor the total number of timepoints. We further represent the $N$ images as vertices of a graph $\mathcal{G}$, which also contains an additional vertex in the centre -- corresponding to a latent subject-specific template (Figure~\ref{fig:spanning-tree}). The vertices corresponding to timepoints are all connected to the centre, creating a spanning tree with $N$ edges associated with a set of $N$ latent transforms $\{\mathcal{T}\}_{n=1, \ldots, N}$ from the template to each image, inducing directionality on the graph $\mathcal{G}$. These latent transforms need to be invertible such that any pair of timepoints are uniquely related by the composition of two transforms along the edge that connects them: the inverted transform from the first time point to the template and the transform from the template to the second timepoint.

We also consider a set of $K>=N$ (and typically $K>>N$) observed transforms (``registrations'') $\{\mathcal{R}\}_{k=1,\ldots, K}$ between  pairs of images of the graph $\mathcal{G}$. These transforms (shown in red in Figure~\ref{fig:spanning-tree}) are computed with a registration algorithm, such that each $\{\mathcal{R}\}_k$ can be seen as a noisy version of the composition of two of the latent transforms $\{\mathcal{T}\}_n$ (which define the ``true'' underlying deformation) -- one of them inverted. Specifically, the registration $\mathcal{R}_k$ between $I_n$ (reference) and $I_{n'}$ (target) images is simply
\begin{equation}
\label{eq:init_model}
    \mathcal{R}_k = \mathcal{T}_{n'} \circ \mathcal{T}^{-1}_n \circ \mathcal{E}_k,
\end{equation}
where $\mathcal{E}_k$ represents the registration error (``noise'').

\subsection{Probabilistic modelling}
\label{sec:prob_framework}
We assume that, for a given subject, any MRI scan across time is the realisation of a random process that randomly deforms the latent subject-specific template and adds noise to the resulting image:
\begin{equation}
\label{eq:generative_template}
    I_n(\bm{x}) = Z(T_n^{-1}(\bm{x})) + \epsilon(x),
\end{equation}
where $Z$ represents the latent (hidden) template, $\bm{x} \in \Omega$ is the spatial location within the image domain $\Omega$, and $\epsilon(x) \sim \text{Laplace}(0,b_{\epsilon})$ is the random noise on the image intensities. 

For the latent transforms $\{\mathcal{T}_n\}$ and the observed registrations $\{\mathcal{R}_k\}$, we assume a probabilistic model such that we can use Bayesian inference to find a set of transforms that allows us to compute a smooth trajectory across time. 
The probabilistic model relies on the assumption that the observed registrations are conditionally independent, given the latent transforms, i.e., $p(\{\mathcal{R}_k\}) = p( \{\mathcal{T}_n\}) \prod_k p(\mathcal{R}_k|\{\mathcal{T}_n\})$. In general, the likelihood of each registration $\mathcal{R}_k$ is parameterised by a set of parameters $\bm{\theta}$ that shape the probabilistic function that we assume fixed (more details in Section~\ref{sec:inference_algorithm}. 
Under these assumptions, the joint probability distribution of the latent transforms, the observed registrations and the likelihood parameters is:
\begin{align}
\label{eq:prob_model}
    p(\{\mathcal{T}_n\},  \{\mathcal{R}_k\}, \bm{\theta}) 
    & = p(\{\mathcal{T}_n\}) p(\bm{\theta}) \prod_{k=1}^K p(\mathcal{R}_k | \{\mathcal{T}_n\}, \bm{\theta})  \nonumber \\
    & = p(\{\mathcal{T}_n\}) \prod_{k=1}^K p(\mathcal{R}_k | \{\mathcal{T}_n\}, \bm{\theta}) 
\end{align}

In practice, the term $\log p(\{\mathcal{T}_n\})$ can be seen as a regulariser on the latent transforms. 

\subsection{Simplification with Lie algebra}
\label{sec:lie_algebra}


The likelihood model~\ref{eq:prob_model} is greatly simplified using deformation models that could be parameterised in the Lie algebra space, including linear and non-linear transforms. 

Let us define $\{\bm{R}_k\}$ and $\{\bm{T}_n\}$ as the log-domain parameterisations whose exponential maps result in the corresponding transformations  $\mathcal{R}_k=\exp\left[\bm{R}_k \right]$ and $\mathcal{T}_k=\exp\left[\bm{T}_k\right]$. The exponential map computation depends on the choice of the deformation model (rigid or nonlinear; see Section~\ref{sec:model_inst} below). Two relevant properties of  Lie algebra are specially useful in this framework. First, the inverse transform is exactly equivalent to its negation in the log-space domain:
\begin{equation*}
  \mathcal{T}_n^{-1} = \exp{\left[-\bm{T}_n\right]}.
\end{equation*}
Second, the composition of transforms can be approximated by the addition after truncating the Baker-Campbell-Hausdorff series at its first term \cite{vercauteren2008symmetric}:
\begin{equation*}
  \mathcal{T}_n \circ \mathcal{T}_{n'} \approx \exp{\left[\bm{T}_n + \bm{T}^T_{n'}\right]};
\end{equation*} 
we note that the formula is actually exact for rigid transforms (more on Section~\ref{sec:model_inst} below). These two properties enable us to linearise the probabilistic model in \ref{eq:init_model} by simply computing the log-space parameterisation:
\begin{equation}
    \label{eq:generative_model}
    \bm{R} = \bm{W} \bm{T} + \bm{\zeta}
\end{equation}
where $\bm{\zeta}$ is the registration error in the log-domain and $\bm{W}$ is the sparse matrix that encodes the path that any $\mathcal{R}$ traverses through the spanning tree (i.e., $\{\mathcal{T}\}_n$).
Hence, for a given spatial transform $\mathcal{R}_k$ between $I_n$ (reference) and $I_{n'}$ (target) images, the $k$-th row of $\bm{W}$ is non-zero at entries $W_{kn}=-1$ and $W_{kn'}=1$. We also assume conditional independence across spatial location, $\bm{x}$, and coordinates $j=1, \ldots, dim(\Omega)=3$. As a result, the likelihood function can be written in terms of the log-space parameterisations as
\begin{equation}
\label{eq:likelihood_function_log_domain}
    p(\mathcal{R}_k | \{\mathcal{T}_n\}, \bm{\theta}) = p(\bm{R}_k | \{\bm{T}_n\}, \bm{\theta}) = \prod_{j=1}^{3} \prod_{\bm{x} \in \Omega}p(\bm{R}^j_k(\bm{x}) | \{\bm{T}^j_n(\bm{x})\}, \bm{\theta}),
\end{equation}
for which we use the Laplace distribution
\begin{equation}
\label{eq:nonrigid_laplace_likelihood}
    \bm{R}_k^j(\bm{x}) \sim \text{Laplace}\left(\bm{W}\bm{T}^j(\bm{x}), b_T\right),
\end{equation}
where $b_T$ is the scale of the Laplace distribution and considered the same for all observations, spatial locations and coordinates. The Laplace distribution has the advantage of being robust against registration errors, as shown in \cite{casamitjana2022robust}. 

Similar to \citep{wu2012feature}, we limit the 
``global drift'' across timepoints to estimate a subject space that lies on the centre of all timepoints. Hence, as the prior distribution, we assume that the composition of the latent transforms (approximated by their sum in the log-domain) follows a Laplace distribution centred at zero:
\begin{equation}
   \label{eq:laplace_prior}
    \left(\sum_{n=0}^{N-1} \bm{T}^j_n(\bm{x}) \right) \sim \text{Laplace}\left(0, b_Z\right),
\end{equation}
where $b_Z$ is the scale of the Laplace distribution, assumed constant for all spatial locations and coordinates. In practice, $b_Z$ is large, as the goal of this prior is make the solution unambiguous and centre it at zero -- rather than strongly regularising the latent transforms. The model parameters are thus $\bm{\theta} = \{b_T, b_Z\}$.

\subsection{Model instantiation}
\label{sec:model_inst}
We present two  deformation models that we apply sequentially: first, a rigid transformation, that creates a shared space $\Omega$ on which the template node is defined, and then a nonlinear transformation, which assumes that all timepoints are resampled on $\Omega$. The final latent transforms are computed as the composition of the rigid and nonlinear components. 


\subsubsection{Rigid transforms}
\label{sec:model_rigid}
Herein this section, let $\{\mathcal{R}_k\}$ and $\{\mathcal{T}_n\}$ be the rigid transforms represented by $4\times 4$ matrices as:
$$
\left(
    \begin{array}{cc}
     \bm{U} & \bm{t} \\
     \bm{0}_{1x3} & 1
\end{array} \right),
$$
where $\bm{U}$ is the $3x3$ rotation matrix and $\bm{t}$ the $3-$dim translation vector. 
The group of rigid transformations in $\mathbb{R}^3$ constitute the special euclidean group $\mathcal{SE}(3)$ and can be parameterised in the log-space domain using a 6-dimensional vector $(\bm{q}, \bm{d})^{\top}$ with two 3-dimensional components:  $\bm{q}\in \mathbb{R}^3$ and $\bm{d}\in \mathbb{R}^3$, that determine the rotation and translation, respectively \citep{blanco2021tutorial}. Thus, the Lie algebra parameterisations of the transforms are $\bm{R}_k=(\bm{q}(R_k), \bm{d}(R_k))^{\top}$ and $\bm{T}_n=(\bm{q}(T_n), \bm{d}(T_n))^{\top}$. Note that we drop the spatial dimension $\bm{x}$ as the parameters are independent of the location.

To compute the log-space parameters, we use the following expressions \citep{blanco2021tutorial}: 
\begin{align}
\label{eq:rigid_log_space_parameterisation}
    \bm{q} &= \frac{\phi}{2\sin{\phi}}\left(U_{32} - U_{23}, U_{13} - U_{31}, U_{21} - U_{12}\right)^{\top} \nonumber \\
\bm{d} &= \bm{P}^{-1}\bm{t},
\end{align}
where $U_{ij}$ is the matrix value corresponding to the $i$-th row and $j$-th column of matrix $U$, and $(\phi, P)$ can be computed as:
\begin{align*}
    \cos(\phi) &= \frac{1}{2}\left(tr(\bm{U})-1\right) \\
    \bm{P}^{-1} &= \bm{I}_3 + 0.5\bm{Q} + \frac{(1 - \frac{\phi\cos{\phi/2}}{2\sin{\phi/2}})}{\phi^2}\bm{Q}^2,
\end{align*}
where
\begin{align*}
  \bm{Q} &= \left(
    \begin{array}{ccc}
       0  & -q_z & q_y\\
       q_z  & 0 & -q_x\\
       -q_y  & q_x & 0
    \end{array}
  \right) \hspace{2mm} \text{and}
  \hspace{2mm} \bm{q} = [q_x, q_y, q_z]^{\top} 
\end{align*}

To compute the exponential maps, we use the closed form expressions for the Lie group parameters, $\bm{U}$ and $\bm{t}$, in \cite{blanco2021tutorial}: 

\begin{align}
    \bm{U} &= \bm{I}_3 + \frac{\sin{\phi}}{\phi}\bm{Q} + \frac{(1 - \cos{\phi})}{\phi^2}\bm{Q}^2 \\
    \bm{t} &= \left(\bm{I}_3 + \frac{(1 - \cos{\phi})}{\phi^2} \bm{Q} + \frac{\phi - \sin{\phi}}{\phi^3}\bm{Q}^3 \right)     \bm{d}  ,
\end{align}

\subsubsection{Non-rigid diffeomorphisms}
\label{sec:model_nonlinear}
Here, we assume that all images are rigidly aligned and resampled onto the same (discrete) spatial domain $\Omega$, which we refer to as \textit{subject space}. In practice, this subject space is an arbitrarily defined 1 mm isotropic grid (further details in Section~\ref{sec:subject_template}). 

With $\Omega$ fixed, we can parameterise a class of nonlinear diffeomorphisms using the Lie group of stationary velocity fields (SVFs, \cite{arsigny2006log}). Let $\{\bm{R}_k(\bm{x})\}$ and $\{\bm{T}_n(\bm{x})\}$ be the SVF infinitesimal generators in the log-space whose integration results in the corresponding diffeomorphisms $\mathcal{R}_k=\exp\left[\bm{R}_k \right]$ and $\mathcal{T}_k=\exp\left[\bm{T}_k \right]$. The scaling-and-squaring approach is used for fast computation of these exponentials \cite{arsigny2006log}.

\subsection{Inference algorithm}
\label{sec:inference_algorithm}

Following the formulation in Section~\ref{sec:prob_framework}, we use Bayesian inference to compute the most likely set of $N$ transforms $\{\mathcal{T}_n\}$ that generate the pairwise image registrations $\{\mathcal{R}_k\}$. 
In a fully Bayesian formulation, the problem of finding the most likely latent transforms requires marginalisation over the parameters we are not seeking to optimise, in this case $\bm{\theta}=\{b_T, b_Z\}$. However, the relationship between these hyperparameters is assumed to be known, yielding the following optimisation function:
\begin{small}
\begin{align}
    \{\hat{\bm{T}}_n\} & = \argmax_{\{\bm{T}_n\}, \bm{\theta} } p(\{\bm{T}_n\}, \bm{\theta}, \{\bm{R}_k\}) \nonumber \\
    & = \argmax_{\{\bm{T}_n\}, \bm{\theta} }  p(\{\bm{T}_n\}) \prod_{k=1}^K\prod_{j=1}^{3} \prod_{\bm{x}\in \Omega} p(\bm{R}^j_k (\bm{x}) | \{\bm{T}^j_n(\bm{x})\}, \bm{\theta}) \nonumber \\
    & = \argmax_{\{\bm{T}_n\}, \bm{\theta} }  \log  p(\{\bm{T}_n\}) +   \sum_{k=1}^K \sum_{j=1}^{3} \sum_{\bm{x}\in \Omega} \log p(\bm{R}^j_k (\bm{x}) | \{\bm{T}^j_n(\bm{x})\}, \bm{\theta}).  \label{eq:opt_model}
\end{align}
\end{small}

Substituting the Laplacian likelihood and prior from Equations~\ref{eq:nonrigid_laplace_likelihood} and \ref{eq:laplace_prior} into Equation~\ref{eq:opt_model}, we obtain the following objective function:
\begin{align}
    \mathcal{O}_{\ell_1} &=  - 2 * 3 (K+1) |\Omega| \log (2b_T) - 2 * 3 |\Omega| \log (2b_Z) \nonumber \\
    &- \frac{1}{b_Z}  \sum_{j=1}^3 \sum_{\bm{x}\in\Omega}
    | \sum_{n=1}^N \bm{T}_n^{j}(\bm{x})|  \nonumber \\
    &  
     - \frac{1}{b_T} \sum_{j=1}^3\sum_{k=1}^K \sum_{\bm{x}\in\Omega}
    |\bm{R}_k^j(\bm{x})-\sum_{n=1}^N W_{kl} \bm{T}_n^j(\bm{x})|. \label{eq:objectiveL1} 
\end{align}
To keep it general, we retain the dependency on the spatial coordinates and location needed for the nonlinear model; note that we would drop it for the rigid case. 

Rearranging terms and switching signs, the cost function to minimise results as follows:
\begin{align}
C_{\ell_1}\left(\bm{T}  (\bm{x})\right) =& \frac{b_T}{b_Z}\sum_{j=1}^3 \sum_{\bm{x}\in\Omega}
    |\sum_{n=1}^N \bm{T}_n^j(\bm{x})|   \nonumber \\
&  + \sum_{j=1}^3 \sum_{k=1}^K \sum_{\bm{x}\in\Omega}
 |\bm{R}_k^j(\bm{x})-\sum_{n=1}^N \bm{W}_{kn} \bm{T}_n^j(\bm{x})|, \label{eq:problemL1_whole} 
\end{align}
which can be solved one spatial location $\bm{x}$ and  coordinate at a time. Thus, we independently solve
\begin{equation}
C_{\ell_1}\left(\bm{T}^j(\bm{x})\right) = 
    \frac{b_T}{b_Z}|\sum_{n=1}^N \bm{T}_n^j(\bm{x})|    + \sum_{k=1}^K  |\bm{R}_k^j(\bm{x})-\sum_{n=1}^N \bm{W}_{kn},\bm{T}_n^j(\bm{x})|, \label{eq:problemL1} 
\end{equation}
at every $\bm{x}$. After visual inspection of the resulting velocity fields, we empirically set the hyperparameter relationship to 1, i.e., $b_T/b_Z=1$, as a trade-off between smoothness and accuracy. Note that in the case of rigid transforms, we solve all parameters at once and are the same for every spatial location. 

The minimisation of Equation~\ref{eq:problemL1} can be rewritten as a linear program in standard form as follows:
\[\arraycolsep=1.4pt\def\arraystretch{1.5}
\begin{array}{rr@{}ll}
\label{eq:linear_program}
\text{minimize}  & \displaystyle \bm{c}^{T}\bm{\Tilde{y}}    & \\
\text{s. t.}     & \displaystyle \bm{A}_1^{T} & \bm{\Tilde{y}} \leq 0,\\
                 & \displaystyle \bm{A}_2^{T} & \bm{\Tilde{y}} \leq 0,\\ 
                 & \displaystyle \bm{A}_3^{T} & \bm{y} \leq -\bm{R}^j(x),\\
                 & \displaystyle \bm{A}_4^{T} & \bm{y} \leq \bm{R}^j(x),\\

\end{array}
\]
where:
\begin{itemize}
\item $\bm{\Tilde{y}} = [D^j_0(\bm{x}), \bm{y}^{T}]^{T}$, is the $(K+N+1)\times 1$ vector of unknown latent variables, concatenating the deviation associated to the regularisation term, $D^j_0(\bm{x})$, and $\bm{y}$, both defined below.
\item $\bm{y} = [D^j_1(\bm{x}),...,D^j_K(\bm{x}),T^j_1(\bm{x}),..., T^j_{N}(\bm{x})]^T$ is a $(K+N)\times 1$ vector concatenating the K absolute deviations of the model, $D^j_k(\bm{x})$ (defined below), and the latent transforms to estimate, $T^j_n(\bm{x})$. 
\item $\bm{c} = [\bm{1}_{K+1}^T, \bm{0}_N^T]^T$, where $\bm{1}_{K+1}$ and $\bm{0}_N$ are the all-one and all-zero vectors with dimensions $(K+1)\times 1$ and $N \times 1$, respectively. 
\item  $\bm{A}_1 = [-1, \bm{0}_K^T, -\bm{1}^T_N]$ is a $(K+N+1)\times 1$ vector.
\item $\bm{A}_2 =[-1, \bm{0}_K^T, \bm{1}^T_N]$.
\item 
$ \bm{A}_3 = [-\bm{I}_{K}, -\bm{W}]$, where $\bm{I}_{K}$ is the $K \times K$ identity matrix.
\item $\bm{A}_4 = [-\bm{I}_{K}, \bm{W}]$.
\end{itemize}

By using vector $\bm{c}$, this linear program effectively minimises the model deviations and it is equivalent to the problem of minimising $\mathcal{C}_{\ell_1}$ in Equation~\ref{eq:problemL1}. The inequality constraints effectively force the deviations $D^{\xi_j}_k(\bm{x})$ to be positive and equal to:
\begin{align*}
    & D^j_{0}(\bm{x}) = \frac{b_T}{b_Z}|\sum_{n=1}^N\bm{T}_n^j(\bm{x})|, \\
    & D^j_k(\bm{x}) = |\bm{R}_k^j(\bm{x})-\sum_{n=1}^N \bm{W}_{kn} \bm{T}_n^j(\bm{x})|, \hspace{3.5mm} \forall k \in (1, \cdots, K).
\end{align*}

The solution is then simply the second part (last $N$ elements) of the vector $\bm{y}$ and can be obtained using well-established linear programming algorithms, such as HiGHS~\citep{huangfu2018parallelizing} (used here) or interior-point methods~\citep{karmarkar1984new,andersen2000mosek}.

\subsection{Registration algorithms}
\label{sec:reg_algorithms}

\subsubsection{Rigid registration}

The presented model in Section~\ref{sec:model_rigid} and inference algorithm in Section~\ref{sec:inference_algorithm} works with any rigid registration algorithm. Here, we use Procrustes analysis (PA, \cite{goodall1991procrustes}) to speed up the registration step -- given that we need to compute $N\times (N-1)/2$ registrations. PA is a statistical shape analysis method that models images as point sets in a given space and minimises the distance between equivalent pairs of points. As in \cite{iglesias2023ready}, we use the centroids for cortical and subcortical ROIs as points in the Euclidean space and find the rotation and translation that minimises the Euclidean distance between centroids. The solution of this optimisation problem is the singular value decomposition of the point sets centred at the origin. It outputs the rotation matrix ($U$) and translation vector ($t$); the closed form expressions from Equation~\ref{eq:rigid_log_space_parameterisation} are used compute the log-space parameters $\bm{v} = (\bm{q}, \bm{d})^{\top}$. The procedure is described in Algorithm\ref{alg:rigid_PA}.


\begin{algorithm}[h]
\caption{Rigid registration}\label{alg:rigid_PA}
 \DontPrintSemicolon
Given a pair of label maps $S_n$, $S_{n'}$ of images $I_n$ and $I_{n'}$\;
Compute centroids $\bm{C}_R(l)$ and $\bm{C}_T(l)$ for $l=1, \ldots, L$.\;
Compute the translations: \;
\begin{itemize}
    \item $\bm{t}_n = \frac{1}{L}\sum_{l=1}^L \bm{C}_n(l),$
    \item $\bm{t}_{n'} = \frac{1}{L}\sum_{l=1}^L \bm{C}_{n'}(l)$
\end{itemize}
Shift point clouds to the origin: \;
\begin{itemize}
    \item $\hat{\bm{C}}_n(l) = \bm{C}_n(l) - \bm{t}_n$
    \item $\hat{\bm{C}}_{n'}(l) = \bm{C}_{n'}(l) - \bm{t}_{n'}$
\end{itemize}
Compute the rotation matrix: 
\begin{align*}
    \bm{V}, \bm{\Delta}, \bm{S} &= \text{SVD}(\hat{\bm{C}}_n(l)\cdot \hat{\bm{C}}^{\top}_{n'}(l)) \\
    \bm{U} &= \bm{S}\bm{V}^{\top}
\end{align*}
Compute the final translation: $ \bm{t} = \bm{t}_{n'} - \bm{U}\cdot \bm{t}_n$\;

Calculate log-space parameters
$  \bm{v}=\left( 
    \begin{array}{c}
          \bm{q} \\
          \bm{d}
    \end{array}\right)$:\;
\begin{align*}
\bm{q} &= \frac{\phi}{2\sin{\phi}}\left(U_{32} - U_{23}, U_{13} - U_{31}, U_{21} - U_{12}\right)^{\top} \\
\bm{d} &= \bm{P}^{-1}\bm{t}
\end{align*}
where 
\begin{align*}
    \cos(\phi) &= \frac{1}{2}\left(tr(\bm{U})-1\right) \\
    \bm{P}^{-1} &= \bm{I}_3 + 0.5\bm{Q} + \frac{(1 - \frac{\phi\cos{\phi/2}}{2\sin{\phi/2}})}{\phi^2}\bm{Q}^2
\end{align*}
\end{algorithm}


\subsubsection{Non-rigid diffeomorphisms}
\label{sec:nonlinear_reg_alg}
An hybrid registration strategy is used to compute the observed SVF maps $\{\bm{R}(\bm{x})\}_k$. For a given $k$-th pair of reference and target images $(I_n, I_{n'})$, we first use SynthMorph \citep{hoffmann2022synthmorph} to compute an initial velocity field at half the image resolution $\bm{\psi}(\bm{x})$.  SynthMorph is a learning-based registration framework that has been globally trained to compute pairwise diffeomorphisms for any pair of MRI contrasts. We then refine the SynthMorph initialisation with a classical method implemented on the graphics processing unit (GPU) using gradient descent to optimise the deformation field. We use a local normalised cross-correlation \citep{avants2008symmetric} to compute image similarity and a gradient penalty on the deformation field as cost function. This refinement step adds a negligible cost to the total calculation in terms of run-time or computational resources and is only used when all timepoints follow approximately the same contrast.

To explicitly enforce symmetry and inverse consistency in the calculated deformation fields, we compute an approximate composition of forward $\bm{\psi}_{nn'}(\bm{x}): I_n \rightarrow I_{n'}$ and backward $\bm{\psi}_{n'n}(\bm{x}): I_{n'} \rightarrow I_n$ velocity fields, similarly to \citep{iglesias2023ready}: $\bm{\psi}(\bm{x}) = 0.5\cdot\bm{\psi}_{nn'}(\bm{x}) - 0.5\cdot\bm{\psi}_{n'n}(\bm{x})$.  Finally, a rescaling layer with linear interpolation is used to get a full resolution velocity field; ``scaling and squaring''  \citep{arsigny2006log} is then used to integrate the SVFs and compute the deformation fields $\{\mathcal{R}_k(\bm{x})\}$ that are inverse consistent, up to the precision of the scale and square algorithm. 

\subsection{Subject-specific template}
\label{sec:subject_template}
The template space is defined on a 1 mm$^3$ isotropic grid $\Omega$. After the linear step of the pipeline, we use the rigid latent transforms to resample all available images on real world coordinates and find the cuboid that include all timepoints. Such cuboid determine the template size and we use a diagonal matrix (plus a translation to centre the image at the origin) to characterise the mapping back between real word and image coordinates.

The subject specific template $Z(\bm{x})$ is built after the nonlinear step of the framework and consists of the intensity image, $Z_I(\bm{x})$, a brain mask, $Z_M(\bm{x})$ and a segmentation map, $Z_S(\bm{x})$, if available. In our case, we use SynthSeg \citep{billot2023robust} to produce brain segmentation maps for all images (which are thresholded to yield brain masks). To build such template, we first align all timepoints (intensities, masks, and one-hot encoded segmentation maps) to the template space using a composition of the rigid and nonlinear deformation fields. Trilinear interpolation from original images is used for resampling. To calculate template intensity image we take the median value of all resampled images as the optimal solution for the Laplacian noise defined in Equation~\ref{eq:generative_template}. For template brain mask, we employ the mean value of all aligned timepoints masks as a measure of brain tissue probability. And segmentation maps are computed as the most likely value after taking the mean on the deformed one-hot encodings of the label volumes.

\section{USLR application examples}
\label{sec:pipeline}

USLR outcomes could serve a variety of applications and downstream tasks. Here, we present 2 different applications: first, an estimation of a single stationary velocity field that characterises the time course of the subject and that could be used, for example, in tensor-based population analyses; and second, a label fusion approach for image segmentation with longitudinal constraints. Note that no assumption on the original image resolution nor the contrast is made. Any resampling made throughout the pipeline is always computed from the original images (i.e., concatenating transforms) to avoid resampling biases due to  smoothing.

\subsection{Stationary subject-specific longitudinal trajectory}
\label{sec:subject_traj}

The composition of the estimated nonlinear latent transforms yields a subject-specific longitudinal trajectory that is varying across  time and not defined outside the followup period of each subject. The purpose of this step is to estimate a single stationary longitudinal trajectory, $\hat{\mathcal{T}}$, as the exponential map of a linear fit on the SVF parameterisation of latent transforms, $\bm{\hat{T}}$:
\begin{equation}
    \label{eq:subject_trajectory}
    \hat{\mathcal{T}}(\bm{x}, t) = exp(t\cdot\bm{\hat{T}}(\bm{x})).
\end{equation}

For that, we use a voxelwise linear model on the SVF maps for each spatial direction and location as follows:
\begin{equation}
\bm{T}^j_n(\bm{x}) = \bm{c}^j(\bm{x}) + \bm{\hat{T}}^j(\bm{x})\cdot t_n, \quad \text{with}\ j\in\{x, y, z\}.
\end{equation}
where $t_n$ is the time to baseline of the $n$-th timepoint and $\bm{c}(\bm{x})$ is the constant map that shifts the origin to the template image.

As in Section~\ref{sec:model_nonlinear}, we assume independence across coordinates and spatial locations. Despite this fact, we observe that smooth latent deformations $\{\mathcal{T}(\bm{x})\}_n$ lead to a smooth subject-specific longitudinal trajectory $\hat{\mathcal{T}}(\bm{x}, t)$. We emphasise that using a linear model on the SVF maps does not lead to a linear model on the deformations due to the exponential relation, as seen in Eq.~\ref{eq:subject_trajectory}. 

This approximation of the subject-specific trajectory can be used, for example, for linear prediction of future timepoints by deforming the nonlinear template at a given time. It constitutes a longitudinal signature of brain anatomical changes for each subject that could be used for population studies, such as tensor- and deformation-based morphometry (TBM/DBM). In such scenario, it is necessary that all deformations lie on the same shared space (e.g., MNI). To align longitudinal signatures of multiple subjects, we use the Pole Ladder introduced in \cite{lorenzi2014efficient} for parallel transport of vector-valued quantities. In short, we transport the approximated subject-specific SVF, defined on subject space as $\hat{\bm{T}}(\bm{x})$, along the deformation field that aligns the subject-specific template with the population template. To normalise each template to the standard space, we use the same registration algorithm as in \ref{sec:nonlinear_reg_alg}

\subsection{Longitudinal segmentation}
Capitalising on the $N$ latent transforms (rigid and non-rigid) that connect all images through the spanning tree defined in Section~\ref{sec:slr_preliminaries}, we compute a time-consistent segmentation on each timepoint space employing the label fusion strategy in \cite{sabuncu2010generative} on the deformed cross-sectional labelmaps. These initial segmentations could be computed using any manual, automatic and semi-automatic method or a combination of those, as long as all timepoints follow the same delineation protocol. Here, we use only timepoints with available T1w images, that are corrected for intensity inhomogeneities and normalised such that the mean value of white matter voxels is 110.

In short, for any given observation defined as reference, we linearly resample both the T1w image and the one-hot encoded labelmaps of the $N-1$ remaining timepoints to the reference space using the computed latent transforms. In our case, we use SynthSeg segmentation for initialisation~\citep{billot2023robust}. Then, we compute the observation's image likelihood using a Gaussian kernel with a standard deviation of 3 on the intensity differences between the reference and the deformed images. Finally, we compute the MAP estimate of the time-consistent segmentation using the image likelihood and the deformed label maps. This step could be seen as the refinement of the cross-sectional segmentation using all subject's observations to introduce consistency across timepoints. 


\section{Experiments and results}
\label{sec:results}
\subsection{Data}
We use the Minimal Interval Resonance Imaging in Alzheimer's Disease (MIRIAD) dataset \citep{malone2013miriad} to exhaustively show the benefits of using the USLR framework for groupwise analysis in longitudinal settings. The MIRIAD dataset is a structural MRI cohort of 46 Alzheimer's disease subjects and 23 elderly healthy aging adults with multiple available T1w scans collected at different time points. Follow up periods per subject range up to two years and sampling points are unevenly spaced from 2 weeks to one year. Moreover, test-retest scans are available at baseline, and 6 and 38 weeks from baseline. We consider two separate subsets from the MIRIAD study: first, we build a small set using the baseline test and retest images as two different timepoints; and second, we build a larger set consisting of all available sessions for all subjects excluding the re-test images. We process all subjects through our USLR processing pipeline as well as through the longitudinal stream of FreeSurfer for comparison with a standard pipeline widely accepted by the community.

Furthermore, an Alzheimer's disease patient from our local dataset is used to better illustrate the benefits of subject-specific analyses. Recruited at the age of 87.5, a total of 11 timepoints across 7.5 years of follow up with its associated T1w scans are readily available and processed through the USLR pipeline.  

\subsection{Longitudinal trajectories: subject-specific analysis}

\begin{figure}[h]
     \centering
     \begin{subfigure}[b]{\columnwidth}
         \centering
         \includegraphics[width=\columnwidth]{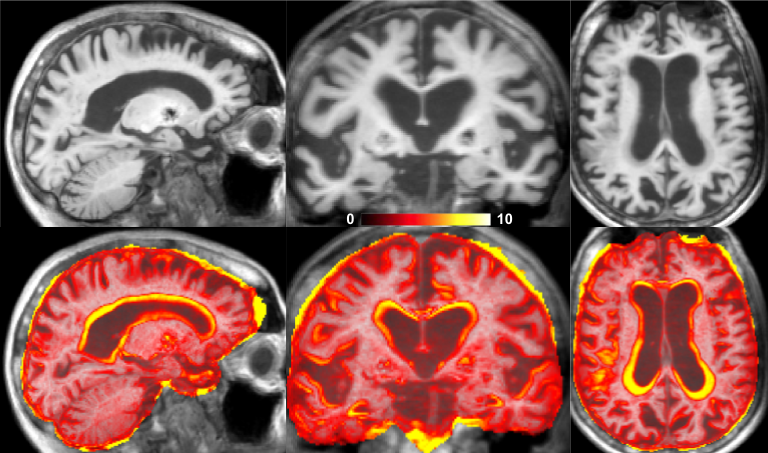}
         \caption{Linear template}
     \end{subfigure}
     \hfill
     \begin{subfigure}[b]{\columnwidth}
         \centering
         \includegraphics[width=\columnwidth]{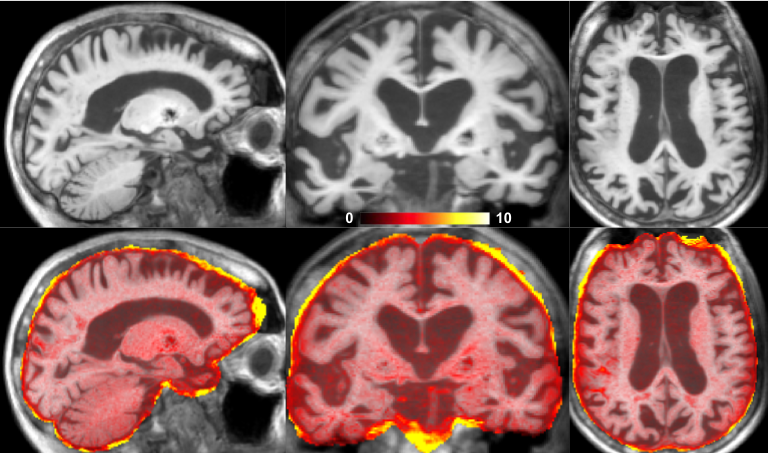}
         \caption{Nonlinear template}
     \end{subfigure}
    \caption{Example of a template of an Alzheimer's disease subject followed-up for a total of 7.5 years with $L=11$ scans from a local cohort. On the top row, the mean T1w subject-specific template and on the bottom row, the standard deviation of the intensities between the template and all the timepoints. The larger variance in the linear template leads to misleading intensity values, such as around the ventricles, where it looks like grey-matter tissue. The nonlinear template is sharper and prevents artificial intensity values.}
    \label{fig:template_std}
\end{figure}

The first and most apparent advantage of the pipeline is the straightforward subject-specific outcomes. On the one hand, a nonlinear template for each subject and unbiased to any timpepoint or atlas is computed. While the benefits of avoiding bias are obvious and discussed in \cite{reuter2012within}, the use of nonlinear transforms yield sharper and more realistic templates compared to linear transforms, as shown in Figure~\ref{fig:template_std}. A linear template is typically used in the standard longitudinal frameworks, such as in Freesurfer. In our framework, the nonlinear stream aims at modelling atrophy and local deformation and adds-up to the initial linear alignment. The resulting better alignment between the template and timepoints reduces the intensity differences among them and avoids artificial intensity values in the image.

\begin{figure}[h]
    \centering
    \includegraphics[width=\columnwidth]{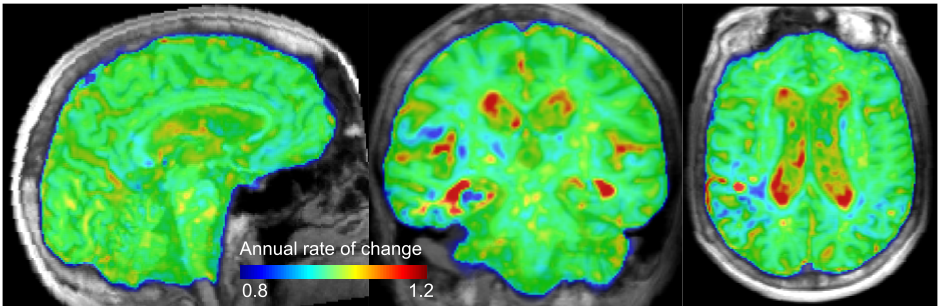}
    \caption{Jacobian determinant of the 1-year subject-specific estimated trajectory of an Alzheimer's disease patient overlaid on its nonlinear template. Hot colours ($>1$) indicate expansion and cold colours ($<1$) indicate contraction over time. The value of each voxel indicates the annual rate of change.}
    \label{fig:jac_AD}
\end{figure}

On the other hand, another direct outcome of the pipeline is the subject-specific trajectory parameterised using stationary velocity fields, as explained in Section~\ref{sec:subject_traj}. Such trajectory could be integrated, using the scaling-squaring algorithm \cite{arsigny2006log}, at any given time to quantify the expected volumetric rate of change. Using tensor based morphometry (TBM), we compute the Jacobian determinant of the deformation to create 3D structural maps of local atrophy. An example of the rate of change over 2 years of an Alzheimer's disease patient is presented in Figure~\ref{fig:jac_AD}, showing the typical expansion around the ventricles and shrinkage of some grey matter regions in the cortex or temporal lobe.

\begin{figure*}[h]
    \centering
    \includegraphics[width=\textwidth]{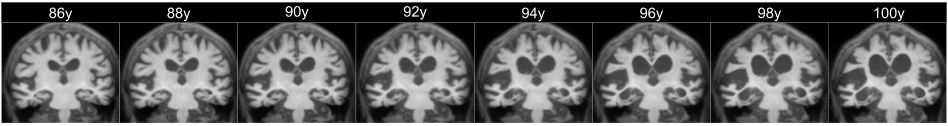}
    \caption{Subject-specific prediction at the voxel level, computed by deforming the nonlinear subject-specific template using the estimated SVF trajectories. The age range shown spans from 1.5 years prior the baseline observation to 5 years after the last observation. The follow up range is [87.5-95] years.}
    \label{fig:lin_pred}
\end{figure*}

Finally, the resulting trajectories could be used for prediction. Due to the SVF parameterisation, the trajectories could be easily inverted to compute forward and backward deformations. Thus, we could interpolate between observations (e.g. missing timepoints) and extrapolate outside the follow up period (prediction) with no blurring. An example of subject-specific evolution that expands the observed period both prior to the baseline image after the last timepoint is shown in Figure~\ref{fig:lin_pred}

\subsection{Longitudinal trajectories: groupwise analysis}
\begin{figure}[h] 
    \centering
    \includegraphics[width=\columnwidth]{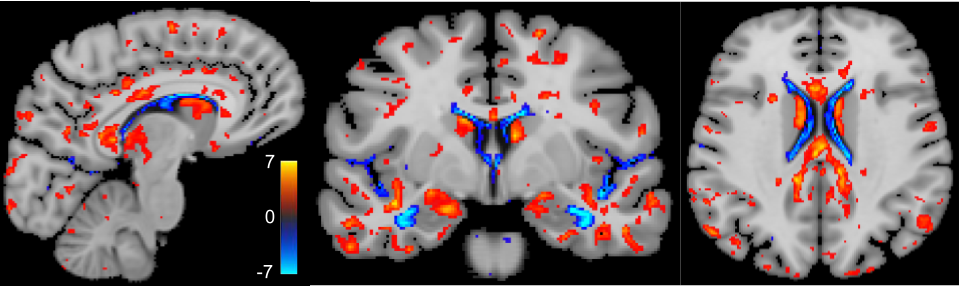}
    \caption{T-test results on the Jacobian determinants between healthy controls and Alzheimer's disease subjects. We show the t-values thresholded at $p=0.05$ corrected for multiple comparisons using FDR.}
    \label{fig:jac_group}
\end{figure}

We now investigate statistical groupwise differences on the longitudinal trajectories between Alzheimer's disease subjects and aging adults without cognitive deficits from the MIRIAD dataset. We first compare the Jacobian determinant maps resampled in a common template, namely the MNI2009a nonlinear template, using voxelwise two-sample t-test. The null hypothesis is that there is no difference on the rate of change between groups. We correct for multiple comparisons using a false discovery rate (FDR) strategy with a corrected p-value threshold at $0.05$ for statistical significance. In Figure~\ref{fig:jac_group} we show the thresholded t-value maps overlaid on the MNI2009a template. We see significant positive and negative differences that can be interpreted as volumetric differences. In particular, increased volume in the area of the ventricles and reduced volume in widespread gray matter regions. 

\begin{figure}[h]
    \centering
    \includegraphics[width=\columnwidth]{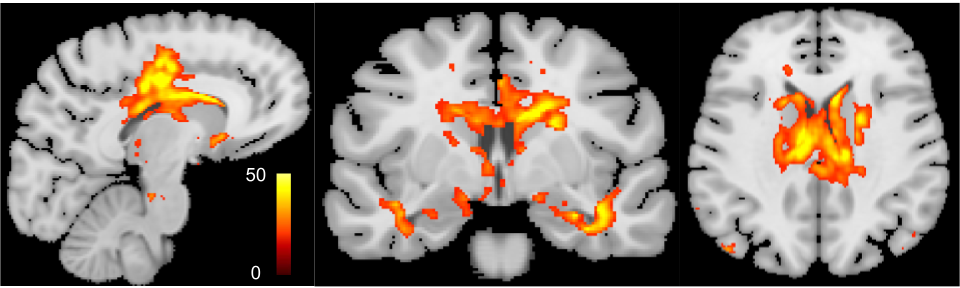}
    \caption{Hotelling T$^2$-test results on the 1 year trajectories between healthy controls and Alzheimer's disease subjects. Values are thresholded at $p=0.05$ corrected for multiple comparisons using FDR.}
    \label{fig:svf_group}
\end{figure}

One of the benefits of the pipeline is that we could test hypotheses directly on the deformations, i.e., deformation based morphometry (DBM). For that, we transport the subject-specific SVFs to MNI space as described in Section~\ref{sec:subject_traj}. On template space, we compute the deformations over a year by integrating the normalised SVFs from $t=0$ to $t=1$. Assuming that there is no difference on the deformations between groups, a multivariate Hotelling's T$^2$-test is employed to check the validity of such null-hypothesis. In Figure~\ref{fig:svf_group} we show the thresholded t-value maps overlaid on the MNI2009a template. We note that typical TBM/DBM studies are cross-sectional and compare the deformation of each group to the template (e.g., \cite{hua2008tensor} ); differently, we compare the longitudinal rate of change between groups once the subject-specific template differences have been minimised through parallel transport to a common template, as previously done in \citep{hadj2016longitudinal}

\subsection{Longitudinal segmentation: groupwise analysis}
Finally, we illustrate the benefits of using the framework for longitudinal segmentation purposes. Since there is no ground-truth dataset with longitudinal segmentations to compare with, we study the test-retest reliability, the sensitivity in detecting changes between groups, the statistical significance of the volumetric trajectories, and its potential impact in study or trial sample sizes.

\begin{figure}[h]
    \centering
    \includegraphics[width=\columnwidth]{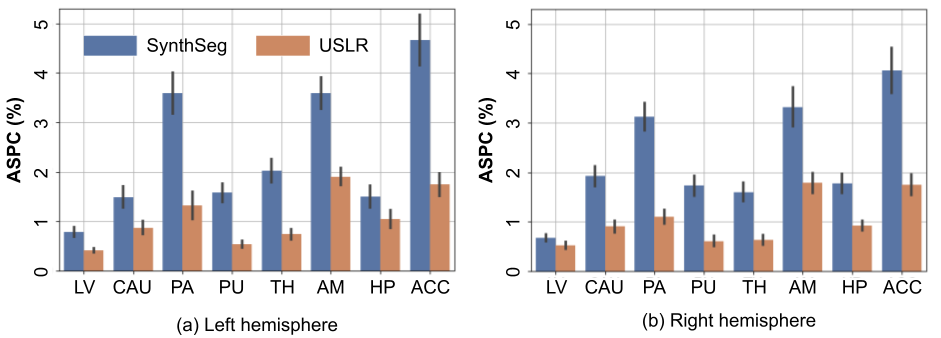}
    \caption{Absolute symmetrised percent change between test-retest images. We compare cross-sectional SynthSeg segmentations in blue with longitudinal USLR in dark orange for subcortical regions on (a) he left hemisphere and (b) the right hemisphere. Each bar represents the median of all subjects and the error bar represents the 95\% confident interval.}
    \label{fig:test-retest}
\end{figure}

Using the test-retest set, we measure the within session variability, which should ideally be zero. We treat the two acquisitions at baseline as separate timepoints and process them through USLR including the longitudinal segmentation step. We then compute the absolute symmetrised percent change (ASPC) as follows,

\begin{equation}
    \text{ASPC} = 100\frac{2|V_2-V_1|}{V_1+V_2},
\end{equation}
and compare the refined longitudinal segmentation against the original cross-sectional SynthSeg segmentations. As shown in Figure~\ref{fig:test-retest}, the longitudinal processing improves the within session reliability reducing potential undesired confounds.

\begin{figure}[h]
    \centering
    \includegraphics[width=\columnwidth]{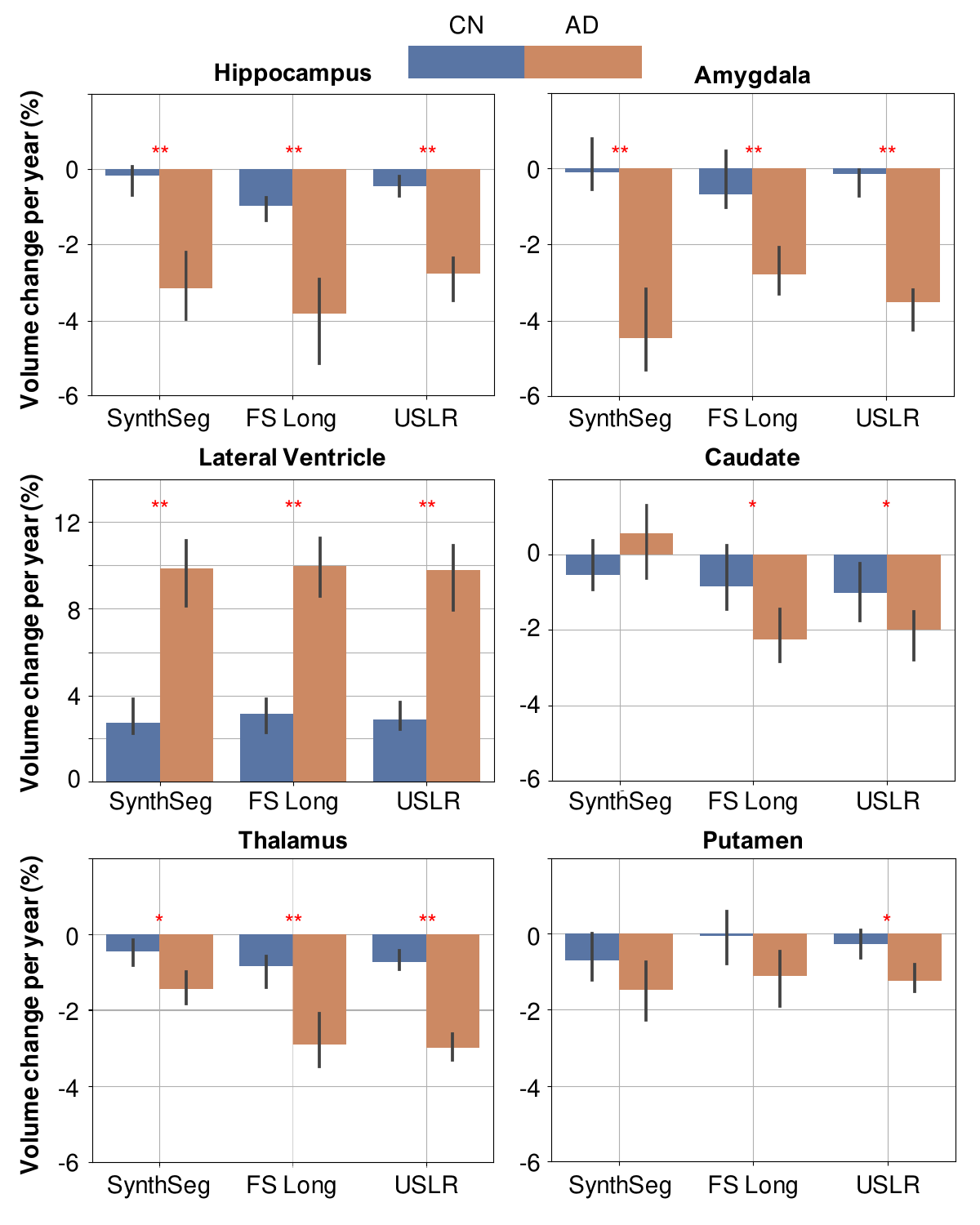}
    \caption{Sensitivity analysis computing the trajectory's slope of 6 different ROI volumes per subject. In each figure, we compare three different segmentation methods: cross-sectional SynthSeg (left), the longitudinal stream of Freesurfer (middle) and our USLR framework (right). Cognitively normal subjects are grouped in blue while AD subjects in dark orange. Each bar shows the median value and the error bars the 95\% confidence interval. Significant differences are found in a Wilcoxon-rank test between groups for (*) $p<5\cdot 10^{-2}$ and (**) $p<1\cdot 10^{-3}$ thresholds.}
    \label{fig:sensitivity-analysis}
\end{figure}

Nonetheless, a very naive algorithm that outputs always the mean value between timepoints would have ideal $ASPC$ despite not being able to capture changes between acquisitions. Thus, we also study the sensitivity of the framework in detecting changes between group's trajectories. We compare USLR with the original SynthSeg and Freesurfer longitudinal segmentations using the entire MIRIAD dataset. In Figure~\ref{fig:sensitivity-analysis}, we show the yearly rate of change with respect the baseline volume for different subcortical regions computed as the mean value between hemispheres. For a more robust estimate, we use the prediction of the linear fit for each subject at $t=0$ as the baseline volume  instead of the value of its first timepoint. Statistical significance is computed  using the Wilcoxon-rank test between groups ('*' stands for $p<0.05$ and '**' stands for $p<0.001$). Clearly, the reduced inter-subject variability of longitudinal processing improves the discrimination power between groups, sometimes at the very mild cost of reduced volumetric change per year. Moreover, we also see less variability between atrophy rates from USLR compared to Freesurfer, specially for the control group.


\begin{figure}[h]
    \centering
    \includegraphics[width=\columnwidth]{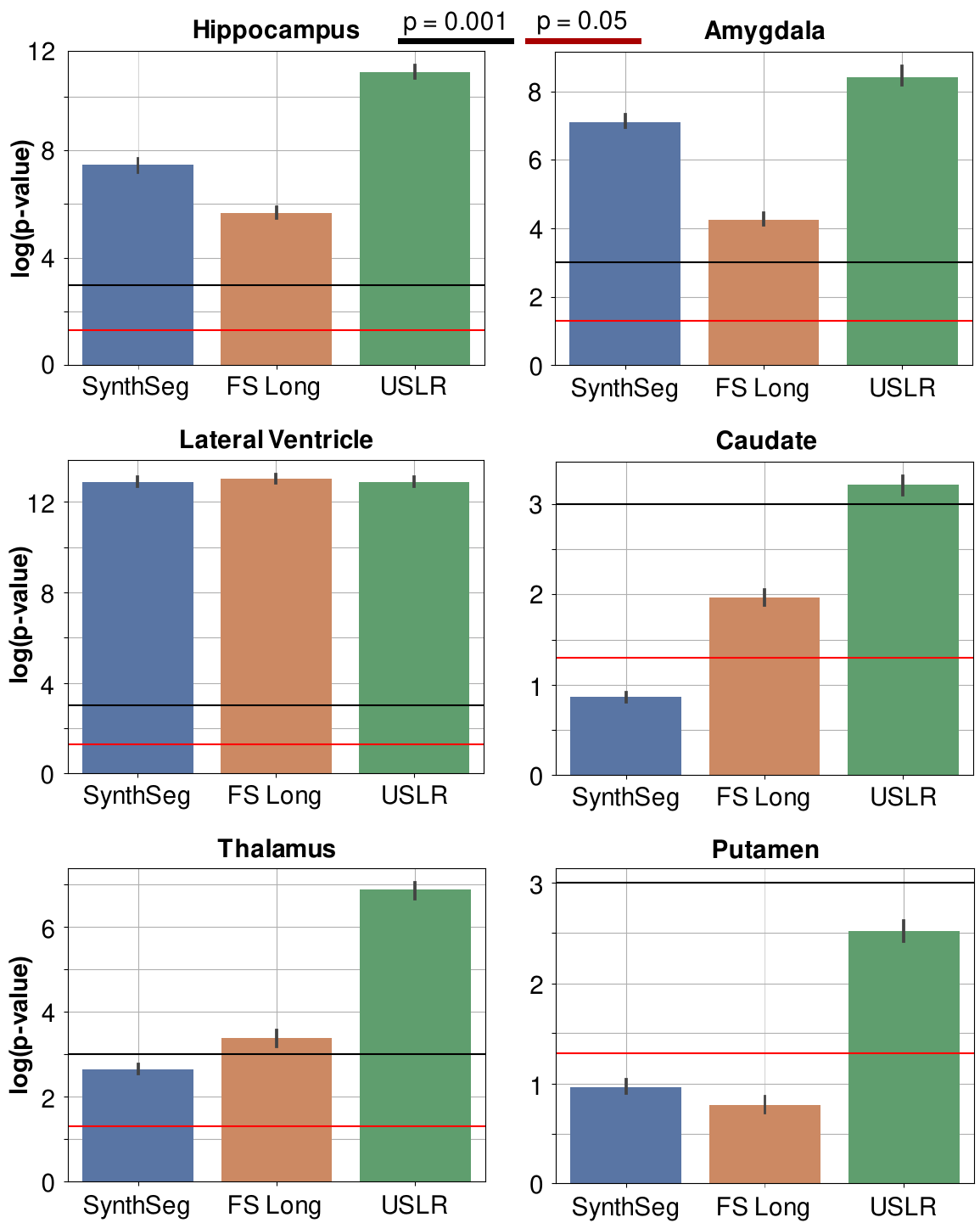}
    \caption{Linear mixed-effects model with random intercept and slope. We plot the $log($p-value$)$ of a contrast comparing time evolution between cognitively normal subjects and AD subjects. The bars represent the median value of $N=1000$ bootstrap sample. We compare three different segmentation methods: cross-sectional SynthSeg (blue), the longitudinal stream of Freesurfer (dark orange) and our USLR framework (green). Red line represents a p-value of $5\cdot 10^{-2}$ and the black line a p-value of $1\cdot 10^{-3}$.}
    \label{fig:lme}
\end{figure}

Moving forward from subject-specific models to an population model, we test the statistical significance of the volumetric trajectories in describing the observations. We use a linear mixed-effects model including seven fixed effects (constant, time from baseline, age, sex, intracranial volume, diagnostic category and interaction between diagnostic category and time from baseline) and random intercept and slope. We employ a contrast on the interaction between diagnosis and time and compute the result on a bootstrap sample (N=1000). In Figure~\ref{fig:lme}, we selected the same subcortical regions and plot the resulting $p$-values with two different thresholds. The USLR volumetric trajectories appear more relevant than both SynthSeg and Freesurfer longitudinal segmentations consistently for almost all subcortical regions.

These results indicate the suitability of the framework on detecting subtle atrophy levels in longitudinal studies which may impact the power of the study or the required sample size. The latter could be easily computed with Eq~\ref{eq:sample_size}. Following \cite{diggle2002analysis} and \cite{reuter2012within} nomenclature,

\begin{equation}
    \label{eq:sample_size}
    m = \frac{2(z_{\alpha} + z_{1-P})^2\sigma^2(1-\rho)}{Ns_x^2d^2}
\end{equation}
we can compute the minimum sample size using different processing techniques. Here, $\sigma$ is the unexplained standard deviation of the observations, $\rho$ is the correlation of repeated observations, $d$ is the target effect size, $N$ is the number of timepoints per subject, $s_x$ the within-subject variance of the variable of interest (in this case, the time between acquisitions and constant across subjects), $P$ the target power of the test, $\alpha$ is the assumed type I error rate, and $z_q$ is the $q$-th quantile of a Gaussian distribution. Most of these are design parameters or constant for a given dataset. Interestingly, only $\sigma$ and $\rho$ differ between processing algorithms. Hence, the sample size reduction when using USLR compared to cross-sectional observations can be expressed as follows:

\begin{equation}
    R = 100\frac{m_{USLR}}{m_{SS}} = 100\frac{\sigma_{USLR}^2(1-\rho_{USLR})}{\sigma_{SS}^2(1-\rho_{SS})}
\end{equation}

We use the baseline and $<2$ weeks observations to compute  volume correlation and inter-subject variability. The test-retest data is not useful here as it scans subjects within the same session limiting the number of factors that explain inter-subject variability, such as hours of sleep or hydration.

\begin{figure}[h]
    \centering
    \includegraphics[width=\columnwidth]{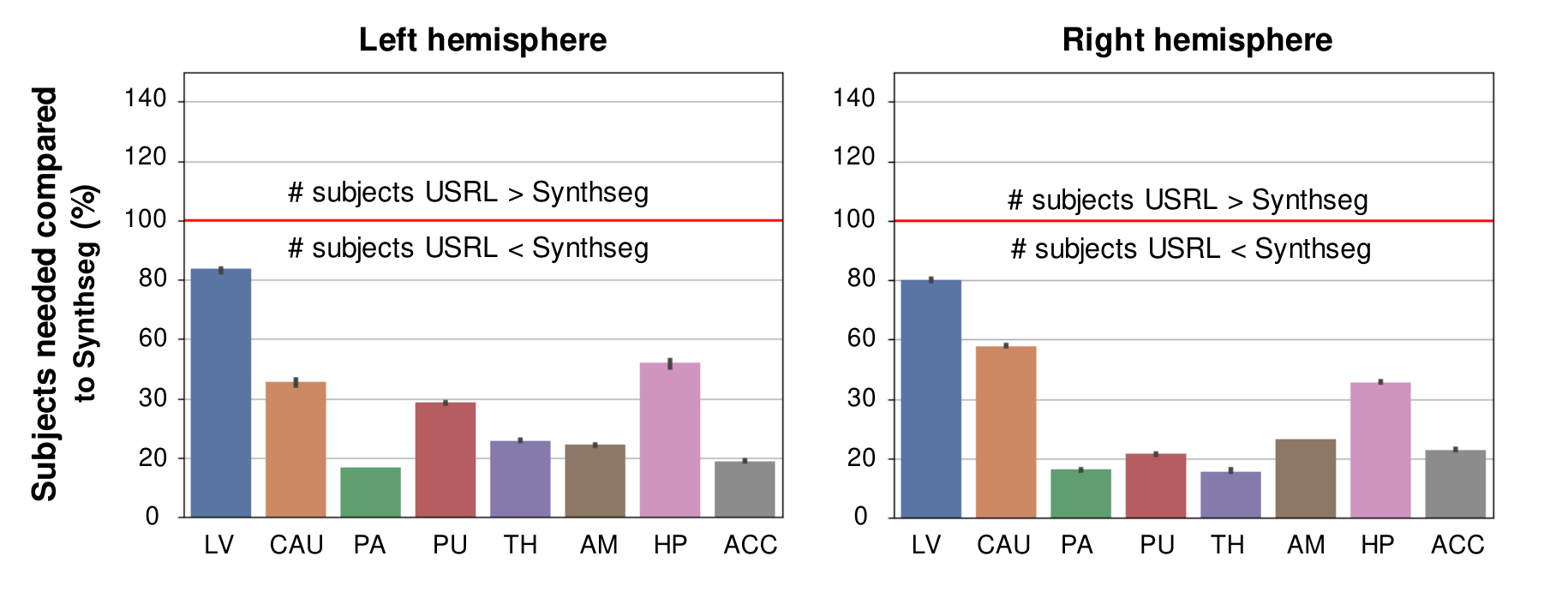}
    \caption{Power analysis showing the reduction in sample size obtained with longitudinal processing of timepoints instead of cross-sectional. Each bar considers a different subcortical region as primary outcome. We show that, for a given study specifications, only a fraction of subjects is needed when using USLR as compared to SynthSeg cross-sectional segmentations. The red line indicates that the same number of subjects are required.}
    \label{fig:power-analysis}
\end{figure}

In Figure~\ref{fig:power-analysis}, we show the results for different brain regions as target endpoints. For example, in a study looking at right hippocampal differences, one only would need roughly 45\% of subjects required if cross-sectional SynthSeg was used. The reduction is even large for other relevant regions such as the thalamus or the amygdala. We hypothesise that the small reduction in the lateral-ventricle responds to the fact that the cross-sectional segmentations are simpler than in other regions, but sample sizes required using USLR are consistently lower than using SynthSeg.

\section{Discussion and conclusions}
\label{sec:discussion}
In this work, we introduced the USLR methodology, a framework for longitudinal registration of brain MRI scans. It capitalises on Bayesian inference and Lie algebra parameterisations of the spatial transforms to find unobserved deformation fields from each timepoint to a latent, unbiased subject-specific template. Importantly, this framework generalises to a variety of transformation models; here we use a rigid transform, to account for global misalignment, followed by a nonlinear stationary velocity fields that model local geometric differences between timepoints. In both cases, the use of learning-based algorithms that are robust to acquisition differences (e.g., scanners, sequences, contrasts) and that provide fast inferences makes the overall pipeline suitable for large scale datasets. Furthermore, we have shown its benefits on a case-control study as compared to using cross-sectional processing pipelines.

The main limitation of this work is the approximation of the composition of nonlinear deformation fields by the summation of SVFs. The error incurred increases with the magnitude of the deformations, but remained unnoticeable for the time span studied in this work. In our experience, it can handle typical follow up times present in clinical trials and observational studies.

We believe that this work serves as a proof-of-concept and that it opens up the use of USLR in multiple applications. We plan to wrap the framework together with some other processing steps (inhomogeneity correction, segmentation, or normalisation to a template) and publish a comprehensive, well tested open-source pipeline available to the community. Following the Bayesian rationale in \cite{cerri2023open}, we also plan to improve the label fusion step using nonlinear deformation fields. This pipeline could be also extended to model brain lesions and be used for treatment follow-up.

\section*{Acknowledgement}
Data used in the preparation of this article were obtained from the MIRIAD database. The MIRIAD investigators did not participate in analysis or writing of this report. The MIRIAD dataset is made available through the support of the UK Alzheimer's Society (Grant RF116). The original data collection was funded through an unrestricted educational grant from GlaxoSmithKline (Grant 6GKC).
Adrià Casamitjana received funding from Ministry of Universities and Recovery, Transformation and Resilience Plan, through UPC (Grant No 2021UPC-MS-67573). R.S has received financial support from the Generalitat de Catalunya (2021-SGR00523), the María de Maeztu Unit of Excellence (Institute of Neurosciences, University of Barcelona, CEX2021-001159-M), and the Spanish Ministry of Science and Innovation (PID2020-118386RA-I00/AEI/10.13039/501100011033). KL was supported by the European Union’s Horizon 2020 research and innovation programme, grant n° 848158 (EarlyCause project). Juan Eugenio Iglesias received funding from NIH grants 1RF1MH123195, 1R01AG070988, 1R01EB031114, 1UM1MH130981, 1RF1AG080371, and a grant from the Jack Satter Foundation. This work is supported by ERC Starting Grant 677697.


\bibliographystyle{elsarticle-harv.bst}\biboptions{authoryear}
\bibliography{refs}

\begin{thebibliography}{54}
\expandafter\ifx\csname natexlab\endcsname\relax\def\natexlab#1{#1}\fi
\providecommand{\url}[1]{\texttt{#1}}
\providecommand{\href}[2]{#2}
\providecommand{\path}[1]{#1}
\providecommand{\DOIprefix}{doi:}
\providecommand{\ArXivprefix}{arXiv:}
\providecommand{\URLprefix}{URL: }
\providecommand{\Pubmedprefix}{pmid:}
\providecommand{\doi}[1]{\href{http://dx.doi.org/#1}{\path{#1}}}
\providecommand{\Pubmed}[1]{\href{pmid:#1}{\path{#1}}}
\providecommand{\bibinfo}[2]{#2}
\ifx\xfnm\relax \def\xfnm[#1]{\unskip,\space#1}\fi
\bibitem[{Agier et~al.(2020)Agier, Valette, K{\'e}chichian, Fanton and
  Prost}]{agier2020hubless}
\bibinfo{author}{Agier, R.}, \bibinfo{author}{Valette, S.},
  \bibinfo{author}{K{\'e}chichian, R.}, \bibinfo{author}{Fanton, L.},
  \bibinfo{author}{Prost, R.}, \bibinfo{year}{2020}.
\newblock \bibinfo{title}{Hubless keypoint-based 3d deformable groupwise
  registration}.
\newblock \bibinfo{journal}{Medical image analysis} \bibinfo{volume}{59},
  \bibinfo{pages}{101564}.
\bibitem[{Andersen and Andersen(2000)}]{andersen2000mosek}
\bibinfo{author}{Andersen, E.D.}, \bibinfo{author}{Andersen, K.D.},
  \bibinfo{year}{2000}.
\newblock \bibinfo{title}{The mosek interior point optimizer for linear
  programming: an implementation of the homogeneous algorithm}, in:
  \bibinfo{booktitle}{High performance optimization}.
  \bibinfo{publisher}{Springer}, pp. \bibinfo{pages}{197--232}.
\bibitem[{Andersson et~al.(2007)Andersson, Jenkinson, Smith
  et~al.}]{andersson2007non}
\bibinfo{author}{Andersson, J.L.}, \bibinfo{author}{Jenkinson, M.},
  \bibinfo{author}{Smith, S.}, et~al., \bibinfo{year}{2007}.
\newblock \bibinfo{title}{Non-linear registration, aka spatial normalisation
  fmrib technical report tr07ja2}.
\newblock \bibinfo{journal}{FMRIB Analysis Group of the University of Oxford}
  \bibinfo{volume}{2}, \bibinfo{pages}{e21}.
\bibitem[{Arsigny et~al.(2006)Arsigny, Commowick, Pennec and
  Ayache}]{arsigny2006log}
\bibinfo{author}{Arsigny, V.}, \bibinfo{author}{Commowick, O.},
  \bibinfo{author}{Pennec, X.}, \bibinfo{author}{Ayache, N.},
  \bibinfo{year}{2006}.
\newblock \bibinfo{title}{A log-euclidean framework for statistics on
  diffeomorphisms}, in: \bibinfo{booktitle}{Medical Image Computing and
  Computer-Assisted Intervention--MICCAI 2006: 9th International Conference,
  Copenhagen, Denmark, October 1-6, 2006. Proceedings, Part I 9},
  \bibinfo{organization}{Springer}. pp. \bibinfo{pages}{924--931}.
\bibitem[{Ashburner(2007)}]{ashburner2007fast}
\bibinfo{author}{Ashburner, J.}, \bibinfo{year}{2007}.
\newblock \bibinfo{title}{A fast diffeomorphic image registration algorithm}.
\newblock \bibinfo{journal}{Neuroimage} \bibinfo{volume}{38},
  \bibinfo{pages}{95--113}.
\bibitem[{Aubert-Broche et~al.(2013)Aubert-Broche, Fonov, Garc{\'\i}a-Lorenzo,
  Mouiha, Guizard, Coup{\'e}, Eskildsen and Collins}]{aubert2013new}
\bibinfo{author}{Aubert-Broche, B.}, \bibinfo{author}{Fonov, V.S.},
  \bibinfo{author}{Garc{\'\i}a-Lorenzo, D.}, \bibinfo{author}{Mouiha, A.},
  \bibinfo{author}{Guizard, N.}, \bibinfo{author}{Coup{\'e}, P.},
  \bibinfo{author}{Eskildsen, S.F.}, \bibinfo{author}{Collins, D.L.},
  \bibinfo{year}{2013}.
\newblock \bibinfo{title}{A new method for structural volume analysis of
  longitudinal brain mri data and its application in studying the growth
  trajectories of anatomical brain structures in childhood}.
\newblock \bibinfo{journal}{Neuroimage} \bibinfo{volume}{82},
  \bibinfo{pages}{393--402}.
\bibitem[{Avants et~al.(2008)Avants, Epstein, Grossman and
  Gee}]{avants2008symmetric}
\bibinfo{author}{Avants, B.B.}, \bibinfo{author}{Epstein, C.L.},
  \bibinfo{author}{Grossman, M.}, \bibinfo{author}{Gee, J.C.},
  \bibinfo{year}{2008}.
\newblock \bibinfo{title}{Symmetric diffeomorphic image registration with
  cross-correlation: evaluating automated labeling of elderly and
  neurodegenerative brain}.
\newblock \bibinfo{journal}{Medical image analysis} \bibinfo{volume}{12},
  \bibinfo{pages}{26--41}.
\bibitem[{Baheti et~al.(2021)Baheti, Waldmannstetter, Chakrabarty, Akbari,
  Bilello, Wiestler, Schwarting, Calabrese, Rudie, Abidi
  et~al.}]{baheti2021brain}
\bibinfo{author}{Baheti, B.}, \bibinfo{author}{Waldmannstetter, D.},
  \bibinfo{author}{Chakrabarty, S.}, \bibinfo{author}{Akbari, H.},
  \bibinfo{author}{Bilello, M.}, \bibinfo{author}{Wiestler, B.},
  \bibinfo{author}{Schwarting, J.}, \bibinfo{author}{Calabrese, E.},
  \bibinfo{author}{Rudie, J.}, \bibinfo{author}{Abidi, S.}, et~al.,
  \bibinfo{year}{2021}.
\newblock \bibinfo{title}{The brain tumor sequence registration challenge:
  establishing correspondence between pre-operative and follow-up mri scans of
  diffuse glioma patients}.
\newblock \bibinfo{journal}{arXiv preprint arXiv:2112.06979} .
\bibitem[{Balakrishnan et~al.(2019)Balakrishnan, Zhao, Sabuncu, Guttag and
  Dalca}]{balakrishnan2019voxelmorph}
\bibinfo{author}{Balakrishnan, G.}, \bibinfo{author}{Zhao, A.},
  \bibinfo{author}{Sabuncu, M.R.}, \bibinfo{author}{Guttag, J.},
  \bibinfo{author}{Dalca, A.V.}, \bibinfo{year}{2019}.
\newblock \bibinfo{title}{Voxelmorph: a learning framework for deformable
  medical image registration}.
\newblock \bibinfo{journal}{IEEE transactions on medical imaging}
  \bibinfo{volume}{38}, \bibinfo{pages}{1788--1800}.
\bibitem[{Billot et~al.(2023a)Billot, Greve, Puonti, Thielscher, Van~Leemput,
  Fischl, Dalca, Iglesias et~al.}]{billot2023synthseg}
\bibinfo{author}{Billot, B.}, \bibinfo{author}{Greve, D.N.},
  \bibinfo{author}{Puonti, O.}, \bibinfo{author}{Thielscher, A.},
  \bibinfo{author}{Van~Leemput, K.}, \bibinfo{author}{Fischl, B.},
  \bibinfo{author}{Dalca, A.V.}, \bibinfo{author}{Iglesias, J.E.}, et~al.,
  \bibinfo{year}{2023}a.
\newblock \bibinfo{title}{Synthseg: Segmentation of brain mri scans of any
  contrast and resolution without retraining}.
\newblock \bibinfo{journal}{Medical image analysis} \bibinfo{volume}{86},
  \bibinfo{pages}{102789}.
\bibitem[{Billot et~al.(2023b)Billot, Magdamo, Cheng, Arnold, Das and
  Iglesias}]{billot2023robust}
\bibinfo{author}{Billot, B.}, \bibinfo{author}{Magdamo, C.},
  \bibinfo{author}{Cheng, Y.}, \bibinfo{author}{Arnold, S.E.},
  \bibinfo{author}{Das, S.}, \bibinfo{author}{Iglesias, J.E.},
  \bibinfo{year}{2023}b.
\newblock \bibinfo{title}{Robust machine learning segmentation for large-scale
  analysis of heterogeneous clinical brain mri datasets}.
\newblock \bibinfo{journal}{Proceedings of the National Academy of Sciences}
  \bibinfo{volume}{120}, \bibinfo{pages}{e2216399120}.
\bibitem[{Blanco-Claraco(2021)}]{blanco2021tutorial}
\bibinfo{author}{Blanco-Claraco, J.L.}, \bibinfo{year}{2021}.
\newblock \bibinfo{title}{A tutorial on se(3) transformation parameterizations
  and on-manifold optimization}.
\newblock \bibinfo{journal}{arXiv preprint arXiv:2103.15980} .
\bibitem[{Casamitjana et~al.(2022)Casamitjana, Lorenzi, Ferraris, Peter, Modat,
  Stevens, Fischl, Vercauteren and Iglesias}]{casamitjana2022robust}
\bibinfo{author}{Casamitjana, A.}, \bibinfo{author}{Lorenzi, M.},
  \bibinfo{author}{Ferraris, S.}, \bibinfo{author}{Peter, L.},
  \bibinfo{author}{Modat, M.}, \bibinfo{author}{Stevens, A.},
  \bibinfo{author}{Fischl, B.}, \bibinfo{author}{Vercauteren, T.},
  \bibinfo{author}{Iglesias, J.E.}, \bibinfo{year}{2022}.
\newblock \bibinfo{title}{Robust joint registration of multiple stains and mri
  for multimodal 3d histology reconstruction: Application to the allen human
  brain atlas}.
\newblock \bibinfo{journal}{Medical image analysis} \bibinfo{volume}{75},
  \bibinfo{pages}{102265}.
\bibitem[{Cerri et~al.(2023)Cerri, Greve, Hoopes, Lundell, Siebner, M{\"u}hlau
  and Van~Leemput}]{cerri2023open}
\bibinfo{author}{Cerri, S.}, \bibinfo{author}{Greve, D.N.},
  \bibinfo{author}{Hoopes, A.}, \bibinfo{author}{Lundell, H.},
  \bibinfo{author}{Siebner, H.R.}, \bibinfo{author}{M{\"u}hlau, M.},
  \bibinfo{author}{Van~Leemput, K.}, \bibinfo{year}{2023}.
\newblock \bibinfo{title}{An open-source tool for longitudinal whole-brain and
  white matter lesion segmentation}.
\newblock \bibinfo{journal}{NeuroImage: Clinical} \bibinfo{volume}{38},
  \bibinfo{pages}{103354}.
\bibitem[{De~Vos et~al.(2019)De~Vos, Berendsen, Viergever, Sokooti, Staring and
  I{\v{s}}gum}]{de2019deep}
\bibinfo{author}{De~Vos, B.D.}, \bibinfo{author}{Berendsen, F.F.},
  \bibinfo{author}{Viergever, M.A.}, \bibinfo{author}{Sokooti, H.},
  \bibinfo{author}{Staring, M.}, \bibinfo{author}{I{\v{s}}gum, I.},
  \bibinfo{year}{2019}.
\newblock \bibinfo{title}{A deep learning framework for unsupervised affine and
  deformable image registration}.
\newblock \bibinfo{journal}{Medical image analysis} \bibinfo{volume}{52},
  \bibinfo{pages}{128--143}.
\bibitem[{Diez et~al.(2014)Diez, Oliver, Cabezas, Valverde, Mart{\'\i},
  Vilanova, Rami{\'o}-Torrent{\`a}, Rovira and Llad{\'o}}]{diez2014intensity}
\bibinfo{author}{Diez, Y.}, \bibinfo{author}{Oliver, A.},
  \bibinfo{author}{Cabezas, M.}, \bibinfo{author}{Valverde, S.},
  \bibinfo{author}{Mart{\'\i}, R.}, \bibinfo{author}{Vilanova, J.C.},
  \bibinfo{author}{Rami{\'o}-Torrent{\`a}, L.}, \bibinfo{author}{Rovira, A.},
  \bibinfo{author}{Llad{\'o}, X.}, \bibinfo{year}{2014}.
\newblock \bibinfo{title}{Intensity based methods for brain mri longitudinal
  registration. a study on multiple sclerosis patients}.
\newblock \bibinfo{journal}{Neuroinformatics} \bibinfo{volume}{12},
  \bibinfo{pages}{365--379}.
\bibitem[{Diggle(2002)}]{diggle2002analysis}
\bibinfo{author}{Diggle, P.}, \bibinfo{year}{2002}.
\newblock \bibinfo{title}{Analysis of longitudinal data}.
\newblock \bibinfo{publisher}{Oxford university press}.
\bibitem[{Dufresne et~al.(2020)Dufresne, Fortun, Kumar, Kremer and
  Noblet}]{dufresne2020joint}
\bibinfo{author}{Dufresne, E.}, \bibinfo{author}{Fortun, D.},
  \bibinfo{author}{Kumar, B.}, \bibinfo{author}{Kremer, S.},
  \bibinfo{author}{Noblet, V.}, \bibinfo{year}{2020}.
\newblock \bibinfo{title}{Joint registration and change detection in
  longitudinal brain mri}, in: \bibinfo{booktitle}{2020 IEEE 17th International
  Symposium on Biomedical Imaging (ISBI)}, \bibinfo{organization}{IEEE}. pp.
  \bibinfo{pages}{104--108}.
\bibitem[{Garcia and Marder(2017)}]{garcia2017statistical}
\bibinfo{author}{Garcia, T.P.}, \bibinfo{author}{Marder, K.},
  \bibinfo{year}{2017}.
\newblock \bibinfo{title}{Statistical approaches to longitudinal data analysis
  in neurodegenerative diseases: Huntington’s disease as a model}.
\newblock \bibinfo{journal}{Current neurology and neuroscience reports}
  \bibinfo{volume}{17}, \bibinfo{pages}{14}.
\bibitem[{Goodall(1991)}]{goodall1991procrustes}
\bibinfo{author}{Goodall, C.}, \bibinfo{year}{1991}.
\newblock \bibinfo{title}{Procrustes methods in the statistical analysis of
  shape}.
\newblock \bibinfo{journal}{Journal of the Royal Statistical Society: Series B
  (Methodological)} \bibinfo{volume}{53}, \bibinfo{pages}{285--321}.
\bibitem[{Hadj-Hamou et~al.(2016)Hadj-Hamou, Lorenzi, Ayache and
  Pennec}]{hadj2016longitudinal}
\bibinfo{author}{Hadj-Hamou, M.}, \bibinfo{author}{Lorenzi, M.},
  \bibinfo{author}{Ayache, N.}, \bibinfo{author}{Pennec, X.},
  \bibinfo{year}{2016}.
\newblock \bibinfo{title}{Longitudinal analysis of image time series with
  diffeomorphic deformations: a computational framework based on stationary
  velocity fields}.
\newblock \bibinfo{journal}{Frontiers in neuroscience} \bibinfo{volume}{10},
  \bibinfo{pages}{236}.
\bibitem[{Hoffmann et~al.(2022)Hoffmann, Billot, Greve, Iglesias, Fischl and
  Dalca}]{hoffmann2022synthmorph}
\bibinfo{author}{Hoffmann, M.}, \bibinfo{author}{Billot, B.},
  \bibinfo{author}{Greve, D.N.}, \bibinfo{author}{Iglesias, J.E.},
  \bibinfo{author}{Fischl, B.}, \bibinfo{author}{Dalca, A.V.},
  \bibinfo{year}{2022}.
\newblock \bibinfo{title}{Synthmorph: learning contrast-invariant registration
  without acquired images}.
\newblock \bibinfo{journal}{IEEE transactions on medical imaging}
  \bibinfo{volume}{41}, \bibinfo{pages}{543--558}.
\bibitem[{Hua et~al.(2008)Hua, Leow, Parikshak, Lee, Chiang, Toga, Jack~Jr,
  Weiner, Thompson, Initiative et~al.}]{hua2008tensor}
\bibinfo{author}{Hua, X.}, \bibinfo{author}{Leow, A.D.},
  \bibinfo{author}{Parikshak, N.}, \bibinfo{author}{Lee, S.},
  \bibinfo{author}{Chiang, M.C.}, \bibinfo{author}{Toga, A.W.},
  \bibinfo{author}{Jack~Jr, C.R.}, \bibinfo{author}{Weiner, M.W.},
  \bibinfo{author}{Thompson, P.M.}, \bibinfo{author}{Initiative, A.D.N.},
  et~al., \bibinfo{year}{2008}.
\newblock \bibinfo{title}{Tensor-based morphometry as a neuroimaging biomarker
  for alzheimer's disease: an mri study of 676 ad, mci, and normal subjects}.
\newblock \bibinfo{journal}{Neuroimage} \bibinfo{volume}{43},
  \bibinfo{pages}{458--469}.
\bibitem[{Huangfu and Hall(2018)}]{huangfu2018parallelizing}
\bibinfo{author}{Huangfu, Q.}, \bibinfo{author}{Hall, J.J.},
  \bibinfo{year}{2018}.
\newblock \bibinfo{title}{Parallelizing the dual revised simplex method}.
\newblock \bibinfo{journal}{Mathematical Programming Computation}
  \bibinfo{volume}{10}, \bibinfo{pages}{119--142}.
\bibitem[{Iglesias(2023)}]{iglesias2023ready}
\bibinfo{author}{Iglesias, J.E.}, \bibinfo{year}{2023}.
\newblock \bibinfo{title}{A ready-to-use machine learning tool for symmetric
  multi-modality registration of brain mri}.
\newblock \bibinfo{journal}{Scientific Reports} \bibinfo{volume}{13},
  \bibinfo{pages}{6657}.
\bibitem[{Iglesias et~al.(2023)Iglesias, Billot, Balbastre, Magdamo, Arnold,
  Das, Edlow, Alexander, Golland and Fischl}]{iglesias2023synthsr}
\bibinfo{author}{Iglesias, J.E.}, \bibinfo{author}{Billot, B.},
  \bibinfo{author}{Balbastre, Y.}, \bibinfo{author}{Magdamo, C.},
  \bibinfo{author}{Arnold, S.E.}, \bibinfo{author}{Das, S.},
  \bibinfo{author}{Edlow, B.L.}, \bibinfo{author}{Alexander, D.C.},
  \bibinfo{author}{Golland, P.}, \bibinfo{author}{Fischl, B.},
  \bibinfo{year}{2023}.
\newblock \bibinfo{title}{Synthsr: A public ai tool to turn heterogeneous
  clinical brain scans into high-resolution t1-weighted images for 3d
  morphometry}.
\newblock \bibinfo{journal}{Science advances} \bibinfo{volume}{9},
  \bibinfo{pages}{eadd3607}.
\bibitem[{Iglesias et~al.(2016)Iglesias, Van~Leemput, Augustinack, Insausti,
  Fischl, Reuter, Initiative et~al.}]{iglesias2016bayesian}
\bibinfo{author}{Iglesias, J.E.}, \bibinfo{author}{Van~Leemput, K.},
  \bibinfo{author}{Augustinack, J.}, \bibinfo{author}{Insausti, R.},
  \bibinfo{author}{Fischl, B.}, \bibinfo{author}{Reuter, M.},
  \bibinfo{author}{Initiative, A.D.N.}, et~al., \bibinfo{year}{2016}.
\newblock \bibinfo{title}{Bayesian longitudinal segmentation of hippocampal
  substructures in brain mri using subject-specific atlases}.
\newblock \bibinfo{journal}{Neuroimage} \bibinfo{volume}{141},
  \bibinfo{pages}{542--555}.
\bibitem[{Joshi et~al.(2004)Joshi, Davis, Jomier and Gerig}]{joshi2004unbiased}
\bibinfo{author}{Joshi, S.}, \bibinfo{author}{Davis, B.},
  \bibinfo{author}{Jomier, M.}, \bibinfo{author}{Gerig, G.},
  \bibinfo{year}{2004}.
\newblock \bibinfo{title}{Unbiased diffeomorphic atlas construction for
  computational anatomy}.
\newblock \bibinfo{journal}{NeuroImage} \bibinfo{volume}{23},
  \bibinfo{pages}{S151--S160}.
\bibitem[{Karch et~al.(2019)Karch, Filevich, Wenger, Lisofsky, Becker, Butler,
  M{\aa}rtensson, Lindenberger, Brandmaier and K{\"u}hn}]{karch2019identifying}
\bibinfo{author}{Karch, J.D.}, \bibinfo{author}{Filevich, E.},
  \bibinfo{author}{Wenger, E.}, \bibinfo{author}{Lisofsky, N.},
  \bibinfo{author}{Becker, M.}, \bibinfo{author}{Butler, O.},
  \bibinfo{author}{M{\aa}rtensson, J.}, \bibinfo{author}{Lindenberger, U.},
  \bibinfo{author}{Brandmaier, A.M.}, \bibinfo{author}{K{\"u}hn, S.},
  \bibinfo{year}{2019}.
\newblock \bibinfo{title}{Identifying predictors of within-person variance in
  mri-based brain volume estimates}.
\newblock \bibinfo{journal}{NeuroImage} \bibinfo{volume}{200},
  \bibinfo{pages}{575--589}.
\bibitem[{Karmarkar(1984)}]{karmarkar1984new}
\bibinfo{author}{Karmarkar, N.}, \bibinfo{year}{1984}.
\newblock \bibinfo{title}{A new polynomial-time algorithm for linear
  programming}, in: \bibinfo{booktitle}{Proceedings of the sixteenth annual ACM
  symposium on Theory of computing}, pp. \bibinfo{pages}{302--311}.
\bibitem[{Klein et~al.(2009)Klein, Staring, Murphy, Viergever and
  Pluim}]{klein2009elastix}
\bibinfo{author}{Klein, S.}, \bibinfo{author}{Staring, M.},
  \bibinfo{author}{Murphy, K.}, \bibinfo{author}{Viergever, M.A.},
  \bibinfo{author}{Pluim, J.P.}, \bibinfo{year}{2009}.
\newblock \bibinfo{title}{Elastix: a toolbox for intensity-based medical image
  registration}.
\newblock \bibinfo{journal}{IEEE transactions on medical imaging}
  \bibinfo{volume}{29}, \bibinfo{pages}{196--205}.
\bibitem[{Kraemer et~al.(2000)Kraemer, Yesavage, Taylor and
  Kupfer}]{kraemer2000can}
\bibinfo{author}{Kraemer, H.C.}, \bibinfo{author}{Yesavage, J.A.},
  \bibinfo{author}{Taylor, J.L.}, \bibinfo{author}{Kupfer, D.},
  \bibinfo{year}{2000}.
\newblock \bibinfo{title}{How can we learn about developmental processes from
  cross-sectional studies, or can we?}
\newblock \bibinfo{journal}{American Journal of Psychiatry}
  \bibinfo{volume}{157}, \bibinfo{pages}{163--171}.
\bibitem[{Lee et~al.(2023)Lee, Alam, Jiang, Cervino, Hu and
  Zhang}]{lee2023seq2morph}
\bibinfo{author}{Lee, D.}, \bibinfo{author}{Alam, S.}, \bibinfo{author}{Jiang,
  J.}, \bibinfo{author}{Cervino, L.}, \bibinfo{author}{Hu, Y.C.},
  \bibinfo{author}{Zhang, P.}, \bibinfo{year}{2023}.
\newblock \bibinfo{title}{Seq2morph: A deep learning deformable image
  registration algorithm for longitudinal imaging studies and adaptive
  radiotherapy}.
\newblock \bibinfo{journal}{Medical Physics} \bibinfo{volume}{50},
  \bibinfo{pages}{970--979}.
\bibitem[{Lee et~al.(2021)Lee, Alam, Jiang, Zhang, Nadeem and
  Hu}]{lee2021deformation}
\bibinfo{author}{Lee, D.}, \bibinfo{author}{Alam, S.R.},
  \bibinfo{author}{Jiang, J.}, \bibinfo{author}{Zhang, P.},
  \bibinfo{author}{Nadeem, S.}, \bibinfo{author}{Hu, Y.c.},
  \bibinfo{year}{2021}.
\newblock \bibinfo{title}{Deformation driven seq2seq longitudinal tumor and
  organs-at-risk prediction for radiotherapy}.
\newblock \bibinfo{journal}{Medical physics} \bibinfo{volume}{48},
  \bibinfo{pages}{4784--4798}.
\bibitem[{Lee et~al.(2019)Lee, Nakamura, Narayanan, Brown, Arnold, Initiative
  et~al.}]{lee2019estimating}
\bibinfo{author}{Lee, H.}, \bibinfo{author}{Nakamura, K.},
  \bibinfo{author}{Narayanan, S.}, \bibinfo{author}{Brown, R.A.},
  \bibinfo{author}{Arnold, D.L.}, \bibinfo{author}{Initiative, A.D.N.}, et~al.,
  \bibinfo{year}{2019}.
\newblock \bibinfo{title}{Estimating and accounting for the effect of mri
  scanner changes on longitudinal whole-brain volume change measurements}.
\newblock \bibinfo{journal}{Neuroimage} \bibinfo{volume}{184},
  \bibinfo{pages}{555--565}.
\bibitem[{Lorenzi et~al.(2013)Lorenzi, Ayache, Frisoni, Pennec, (ADNI
  et~al.}]{lorenzi2013lcc}
\bibinfo{author}{Lorenzi, M.}, \bibinfo{author}{Ayache, N.},
  \bibinfo{author}{Frisoni, G.B.}, \bibinfo{author}{Pennec, X.},
  \bibinfo{author}{(ADNI, A.D.N.I.}, et~al., \bibinfo{year}{2013}.
\newblock \bibinfo{title}{Lcc-demons: a robust and accurate symmetric
  diffeomorphic registration algorithm}.
\newblock \bibinfo{journal}{NeuroImage} \bibinfo{volume}{81},
  \bibinfo{pages}{470--483}.
\bibitem[{Lorenzi and Pennec(2014)}]{lorenzi2014efficient}
\bibinfo{author}{Lorenzi, M.}, \bibinfo{author}{Pennec, X.},
  \bibinfo{year}{2014}.
\newblock \bibinfo{title}{Efficient parallel transport of deformations in time
  series of images: from schild’s to pole ladder}.
\newblock \bibinfo{journal}{Journal of mathematical imaging and vision}
  \bibinfo{volume}{50}, \bibinfo{pages}{5--17}.
\bibitem[{Malone et~al.(2013)Malone, Cash, Ridgway, MacManus, Ourselin, Fox and
  Schott}]{malone2013miriad}
\bibinfo{author}{Malone, I.B.}, \bibinfo{author}{Cash, D.},
  \bibinfo{author}{Ridgway, G.R.}, \bibinfo{author}{MacManus, D.G.},
  \bibinfo{author}{Ourselin, S.}, \bibinfo{author}{Fox, N.C.},
  \bibinfo{author}{Schott, J.M.}, \bibinfo{year}{2013}.
\newblock \bibinfo{title}{Miriad—public release of a multiple time point
  alzheimer's mr imaging dataset}.
\newblock \bibinfo{journal}{NeuroImage} \bibinfo{volume}{70},
  \bibinfo{pages}{33--36}.
\bibitem[{Maxwell and Cole(2007)}]{maxwell2007bias}
\bibinfo{author}{Maxwell, S.E.}, \bibinfo{author}{Cole, D.A.},
  \bibinfo{year}{2007}.
\newblock \bibinfo{title}{Bias in cross-sectional analyses of longitudinal
  mediation.}
\newblock \bibinfo{journal}{Psychological methods} \bibinfo{volume}{12},
  \bibinfo{pages}{23}.
\bibitem[{Modat et~al.(2010)Modat, Ridgway, Taylor, Lehmann, Barnes, Hawkes,
  Fox and Ourselin}]{modat2010fast}
\bibinfo{author}{Modat, M.}, \bibinfo{author}{Ridgway, G.R.},
  \bibinfo{author}{Taylor, Z.A.}, \bibinfo{author}{Lehmann, M.},
  \bibinfo{author}{Barnes, J.}, \bibinfo{author}{Hawkes, D.J.},
  \bibinfo{author}{Fox, N.C.}, \bibinfo{author}{Ourselin, S.},
  \bibinfo{year}{2010}.
\newblock \bibinfo{title}{Fast free-form deformation using graphics processing
  units}.
\newblock \bibinfo{journal}{Computer methods and programs in biomedicine}
  \bibinfo{volume}{98}, \bibinfo{pages}{278--284}.
\bibitem[{Mok and Chung(2022a)}]{mok2022robust}
\bibinfo{author}{Mok, T.C.}, \bibinfo{author}{Chung, A.C.},
  \bibinfo{year}{2022}a.
\newblock \bibinfo{title}{Robust image registration with absent correspondences
  in pre-operative and follow-up brain mri scans of diffuse glioma patients},
  in: \bibinfo{booktitle}{International MICCAI Brainlesion Workshop},
  \bibinfo{organization}{Springer}. pp. \bibinfo{pages}{231--240}.
\bibitem[{Mok and Chung(2022b)}]{mok2022unsupervised}
\bibinfo{author}{Mok, T.C.}, \bibinfo{author}{Chung, A.C.},
  \bibinfo{year}{2022}b.
\newblock \bibinfo{title}{Unsupervised deformable image registration with
  absent correspondences in pre-operative and post-recurrence brain tumor mri
  scans}, in: \bibinfo{booktitle}{International Conference on Medical Image
  Computing and Computer-Assisted Intervention},
  \bibinfo{organization}{Springer}. pp. \bibinfo{pages}{25--35}.
\bibitem[{Morey et~al.(2010)Morey, Selgrade, Wagner, Huettel, Wang and
  McCarthy}]{morey2010scan}
\bibinfo{author}{Morey, R.A.}, \bibinfo{author}{Selgrade, E.S.},
  \bibinfo{author}{Wagner, H.R.}, \bibinfo{author}{Huettel, S.A.},
  \bibinfo{author}{Wang, L.}, \bibinfo{author}{McCarthy, G.},
  \bibinfo{year}{2010}.
\newblock \bibinfo{title}{Scan--rescan reliability of subcortical brain volumes
  derived from automated segmentation}.
\newblock \bibinfo{journal}{Human brain mapping} \bibinfo{volume}{31},
  \bibinfo{pages}{1751--1762}.
\bibitem[{Nyberg et~al.(2010)Nyberg, Salami, Andersson, Eriksson, Kalpouzos,
  Kauppi, Lind, Pudas, Persson and Nilsson}]{nyberg2010longitudinal}
\bibinfo{author}{Nyberg, L.}, \bibinfo{author}{Salami, A.},
  \bibinfo{author}{Andersson, M.}, \bibinfo{author}{Eriksson, J.},
  \bibinfo{author}{Kalpouzos, G.}, \bibinfo{author}{Kauppi, K.},
  \bibinfo{author}{Lind, J.}, \bibinfo{author}{Pudas, S.},
  \bibinfo{author}{Persson, J.}, \bibinfo{author}{Nilsson, L.G.},
  \bibinfo{year}{2010}.
\newblock \bibinfo{title}{Longitudinal evidence for diminished frontal cortex
  function in aging}.
\newblock \bibinfo{journal}{Proceedings of the National Academy of Sciences}
  \bibinfo{volume}{107}, \bibinfo{pages}{22682--22686}.
\bibitem[{Reuter et~al.(2012)Reuter, Schmansky, Rosas and
  Fischl}]{reuter2012within}
\bibinfo{author}{Reuter, M.}, \bibinfo{author}{Schmansky, N.J.},
  \bibinfo{author}{Rosas, H.D.}, \bibinfo{author}{Fischl, B.},
  \bibinfo{year}{2012}.
\newblock \bibinfo{title}{Within-subject template estimation for unbiased
  longitudinal image analysis}.
\newblock \bibinfo{journal}{Neuroimage} \bibinfo{volume}{61},
  \bibinfo{pages}{1402--1418}.
\bibitem[{Sabuncu et~al.(2010)Sabuncu, Yeo, Van~Leemput, Fischl and
  Golland}]{sabuncu2010generative}
\bibinfo{author}{Sabuncu, M.R.}, \bibinfo{author}{Yeo, B.T.},
  \bibinfo{author}{Van~Leemput, K.}, \bibinfo{author}{Fischl, B.},
  \bibinfo{author}{Golland, P.}, \bibinfo{year}{2010}.
\newblock \bibinfo{title}{A generative model for image segmentation based on
  label fusion}.
\newblock \bibinfo{journal}{IEEE transactions on medical imaging}
  \bibinfo{volume}{29}, \bibinfo{pages}{1714--1729}.
\bibitem[{Schnabel et~al.(2001)Schnabel, Rueckert, Quist, Blackall,
  Castellano-Smith, Hartkens, Penney, Hall, Liu, Truwit
  et~al.}]{schnabel2001generic}
\bibinfo{author}{Schnabel, J.A.}, \bibinfo{author}{Rueckert, D.},
  \bibinfo{author}{Quist, M.}, \bibinfo{author}{Blackall, J.M.},
  \bibinfo{author}{Castellano-Smith, A.D.}, \bibinfo{author}{Hartkens, T.},
  \bibinfo{author}{Penney, G.P.}, \bibinfo{author}{Hall, W.A.},
  \bibinfo{author}{Liu, H.}, \bibinfo{author}{Truwit, C.L.}, et~al.,
  \bibinfo{year}{2001}.
\newblock \bibinfo{title}{A generic framework for non-rigid registration based
  on non-uniform multi-level free-form deformations}, in:
  \bibinfo{booktitle}{Medical Image Computing and Computer-Assisted
  Intervention--MICCAI 2001: 4th International Conference Utrecht, The
  Netherlands, October 14--17, 2001 Proceedings 4},
  \bibinfo{organization}{Springer}. pp. \bibinfo{pages}{573--581}.
\bibitem[{Sharma et~al.(2013)Sharma, Rousseau, Heitz, Rumbach and
  Armspach}]{sharma2013estimation}
\bibinfo{author}{Sharma, S.}, \bibinfo{author}{Rousseau, F.},
  \bibinfo{author}{Heitz, F.}, \bibinfo{author}{Rumbach, L.},
  \bibinfo{author}{Armspach, J.P.}, \bibinfo{year}{2013}.
\newblock \bibinfo{title}{On the estimation and correction of bias in local
  atrophy estimations using example atrophy simulations}.
\newblock \bibinfo{journal}{Computerized Medical Imaging and Graphics}
  \bibinfo{volume}{37}, \bibinfo{pages}{538--551}.
\bibitem[{Vercauteren et~al.(2008)Vercauteren, Pennec, Perchant and
  Ayache}]{vercauteren2008symmetric}
\bibinfo{author}{Vercauteren, T.}, \bibinfo{author}{Pennec, X.},
  \bibinfo{author}{Perchant, A.}, \bibinfo{author}{Ayache, N.},
  \bibinfo{year}{2008}.
\newblock \bibinfo{title}{Symmetric log-domain diffeomorphic registration: A
  demons-based approach}, in: \bibinfo{booktitle}{International conference on
  medical image computing and computer-assisted intervention},
  \bibinfo{organization}{Springer}. pp. \bibinfo{pages}{754--761}.
\bibitem[{Wodzinski et~al.(2022)Wodzinski, Jurgas, Marini, Atzori and
  M{\"u}ller}]{wodzinski2022unsupervised}
\bibinfo{author}{Wodzinski, M.}, \bibinfo{author}{Jurgas, A.},
  \bibinfo{author}{Marini, N.}, \bibinfo{author}{Atzori, M.},
  \bibinfo{author}{M{\"u}ller, H.}, \bibinfo{year}{2022}.
\newblock \bibinfo{title}{Unsupervised method for intra-patient registration of
  brain magnetic resonance images based on objective function weighting by
  inverse consistency: Contribution to the brats-reg challenge}, in:
  \bibinfo{booktitle}{International MICCAI Brainlesion Workshop},
  \bibinfo{organization}{Springer}. pp. \bibinfo{pages}{241--251}.
\bibitem[{Wu et~al.(2012)Wu, Wang, Jia and Shen}]{wu2012feature}
\bibinfo{author}{Wu, G.}, \bibinfo{author}{Wang, Q.}, \bibinfo{author}{Jia,
  H.}, \bibinfo{author}{Shen, D.}, \bibinfo{year}{2012}.
\newblock \bibinfo{title}{Feature-based groupwise registration by hierarchical
  anatomical correspondence detection}.
\newblock \bibinfo{journal}{Human brain mapping} \bibinfo{volume}{33},
  \bibinfo{pages}{253--271}.
\bibitem[{Yang et~al.(2017)Yang, Kwitt, Styner and
  Niethammer}]{yang2017quicksilver}
\bibinfo{author}{Yang, X.}, \bibinfo{author}{Kwitt, R.},
  \bibinfo{author}{Styner, M.}, \bibinfo{author}{Niethammer, M.},
  \bibinfo{year}{2017}.
\newblock \bibinfo{title}{Quicksilver: Fast predictive image registration--a
  deep learning approach}.
\newblock \bibinfo{journal}{NeuroImage} \bibinfo{volume}{158},
  \bibinfo{pages}{378--396}.
\bibitem[{Young et~al.(2022)Young, Balbastre, Dalca, Wells, Iglesias and
  Fischl}]{young2022superwarp}
\bibinfo{author}{Young, S.I.}, \bibinfo{author}{Balbastre, Y.},
  \bibinfo{author}{Dalca, A.V.}, \bibinfo{author}{Wells, W.M.},
  \bibinfo{author}{Iglesias, J.E.}, \bibinfo{author}{Fischl, B.},
  \bibinfo{year}{2022}.
\newblock \bibinfo{title}{Superwarp: Supervised learning and warping on u-net
  for invariant subvoxel-precise registration}, in:
  \bibinfo{booktitle}{International Workshop on Biomedical Image Registration},
  \bibinfo{organization}{Springer}. pp. \bibinfo{pages}{103--115}.
\bibitem[{Yushkevich et~al.(2010)Yushkevich, Avants, Das, Pluta, Altinay,
  Craige, Initiative et~al.}]{yushkevich2010bias}
\bibinfo{author}{Yushkevich, P.A.}, \bibinfo{author}{Avants, B.B.},
  \bibinfo{author}{Das, S.R.}, \bibinfo{author}{Pluta, J.},
  \bibinfo{author}{Altinay, M.}, \bibinfo{author}{Craige, C.},
  \bibinfo{author}{Initiative, A.D.N.}, et~al., \bibinfo{year}{2010}.
\newblock \bibinfo{title}{Bias in estimation of hippocampal atrophy using
  deformation-based morphometry arises from asymmetric global normalization: an
  illustration in adni 3 t mri data}.
\newblock \bibinfo{journal}{Neuroimage} \bibinfo{volume}{50},
  \bibinfo{pages}{434--445}.

\end{thebibliography}
\end{document}